\documentclass[prb,preprint,amsmath,amssymb]{revtex4}
\usepackage{graphicx}
\begin{document}
\author{Kevin Leung,$^{1*}$ Susan~B.~Rempe,$^1$ Michael E.~Foster,$^1$
Yuguang Ma,$^2$ Julibeth M.~Martinez del la Hoz,$^2$ Na~Sai,$^3$ and Perla B.~Balbuena$^2$}
\affiliation{$^1$Sandia National Laboratories, MS 1415, Albuquerque, NM 87185\\
$^2$Department of Chemical Engineering, Texas A\&M University, College
Station, TX 77843\\
$^3$Department of Physics, University of Texas at Austin, Austin, TX 78712 \\
$^*${\tt kleung@sandia.gov}}
\date{\today}
\title{Modeling Electrochemical Decomposition of Fluoroethylene Carbonate
on Silicon Anode Surfaces in Lithium Ion Batteries}

\input epsf

\begin{abstract}

Fluoroethylene carbonate (FEC) shows promise as an electrolyte additive
for improving passivating solid-electrolyte interphase (SEI) films
on silicon anodes used in lithium ion batteries (LIB).  We apply
density functional theory (DFT), {\it ab initio} molecular dynamics (AIMD),
and quantum chemistry techniques to examine excess-electron-induced
FEC molecular decomposition mechanisms that lead to FEC-modified SEI.
We consider one- and two-electron reactions using cluster models
and explicit interfaces between liquid electrolyte and model Li$_x$Si$_y$
surfaces, respectively.  FEC is found to exhibit more varied reaction
pathways than unsubstituted ethylene carbonate.  The initial bond-breaking
events and products of one- and two-electron reactions are qualitatively
similar, with a fluoride ion detached in both cases.
However, most one-electron products are charge-neutral, not anionic, and
may not coalesce to form effective Li$^+$-conducting SEI unless they are
further reduced or take part in other reactions.  The implications of these
reactions to silicon-anode based LIB are discussed.

\end{abstract}

\maketitle

\section{Introduction}

Solid-electrolyte interphase (SEI) films that passivate low voltage
anode surfaces are important for lithium ion battery
operations.\cite{book2,book,review,novak_review,intro1}  These films arise
from electrochemical reduction and subsequent breakdown of the organic
solvent-based electrolyte, which is unstable under battery charging potentials.
With the correct choice of electrolyte, stable SEI films are formed, blocking
further electron transfer from the anode to the electrolyte, yet permitting
lithium (Li$^+$) transport.  Small amounts of additives are often added
to the electrolyte to modify the structure and performance of SEI films.

Additives may be particularly important for the next generation of anode
battery materials like silicon (Si), tin, and other alloys.  For example, Si
promises much higher lithium ion capacities than commercially used
graphite anodes.\cite{si_review}  However, Si exhibits large volumetric
expansion during Li intercalation.  This often results in cracking, leading
to detachment of particles from each other\cite{fracture,sullivan} and
perhaps from SEI films formed during previous cycles.  There appears to be a
need to continuously reform the SEI layer after each cycle, leading to
consumption of Li$^+$.   Thus, in general, Si anodes have not been
sufficiently stable
over hundreds of charge/discharge cycles to be useful for commercial automobile
batteries.  SEI films on traditional graphite anodes yield less capacity fade,
even though chemical compositions of SEI films on the two materials appear
similar.\cite{si_sei1,si_sei2,si_sei3,si_sei4,si_sei5,si_sei6,si_sei7,si_sei8}

Recently, fluoroethylene carbonate (FEC, Fig.~\ref{fig1}a) was found to
improve the cycling behavior of silicon electrodes in half cell
configurations.\cite{fec0,fec1,fec2,fec3,fec4,fec5,fec6,fec7,fec8}
Electrochemical impedance spectroscopy revealed that FEC-modified SEI films
on Si yield lower Li$^+$ transfer resistance, while atomic force microscopy
reported smoother SEI surfaces.  Mass spectroscopy and Fourier transform
infrared spectroscopy techniques showed little evidence of C-F
covalently-bonded motifs in the SEI layers.  Beyond these broad
areas of agreement, details of SEI structure and composition reported by
different groups differed.  There are discrepancies regarding whether FEC
increases the film thickness and LiF composition, and regarding whether FEC
and vinylene carbonate (VC) have similar additive effects.
Oxalate\cite{fec1,fec3} and polycarbonate\cite{fec1,fec2,fec3,fec5} SEI
components are reported in some studies, but not others.

Closely related to SEI chemical composition is the chemical mechanisms
responsible for decomposition of FEC.  The sequence of covalent bond
breaking, the propagation of reaction intermediates to form polymeric products
(if any), and the number of electrons transferred to reacting complexes
at each stage have not been elucidated experimentally.\cite{note0,e2,abraham}
The apparent absence of C-F bonds in the SEI has prompted at least one
experimental work to speculate that volatile fluorine-containing hydrocarbon
components like C$_2$H$_3$F are removed as gas products before they can form
SEI films.\cite{japan} Another proposed mechanism involves the elimination of
hydrogen fluoride (HF) from FEC to form vinylene carbonate (VC)
deriviatives.\cite{vc}  Others have disputed this claim, citing observed
differences between FEC- and VC-derived SEI films.\cite{fec3}

Electronic structure modeling is an excellent complement to experimental
analysis.\cite{bal01,bal02,bal02a,han04,e2,bedrov,tasaki2010,gewirth,interface,model0,model1,model2,model3,model4}
While unlikely to yield the entire sequence of reactions and precise product
branching ratios, modeling is useful for interrogating whether
each proposed bond-breaking/making step is exothermic, has a sufficiently
low free energy barrier, or whether it will be superceded by other
reaction pathways.  Concerns about theoretical accuracy can be alleviated
by using multiple density functional theory (DFT) and quantum chemistry
methods to cross-check predictions.  The role of electrode surfaces in
promoting/hindering SEI film formation reactions can also be elucidated.
Many insightful and evocative modeling works on bare crystalline Si
bulk,\cite{model0} surfaces\cite{model1,model2,model3,model4} and their
interactions with lithium metal (as Li$^+$ source) have been published, but
so far they have not focused on SEI formation mechanisms.  The Si models used
in these works also tend to be bare Si surfaces, which can exist only under
ultrahigh vacuum (UHV), not electrochemical, conditions.

Our modeling effort focuses on the initial steps of FEC decomposition,
disputes the HF and CH$_2$CHF reaction pathways proposed in the literature,
and reveals lower barrier reaction mechanisms.  Our work also explores
subsequent SEI-formation steps that can occur.  The two types of model used
here dovetails with the two regimes of SEI film formation widely associated
with EC/dimethyl carbonate (DMC) mixtures on graphite anodes and freshly cut
metal surfaces.\cite{book,review,e2}  The initial step consists of fast,
two-electron attack; the later stage involves one-electron reduction of solvent
or salt molecules.  The latter dominates when electron tunneling/transfer
through the half-formed SEI layer slows down.  This two-stage mechanism
may be consistent with the two-layer SEI structure revealed in experiments.
An inorganic inner layer conducting Li$^+$ but blocking solvent and $e^-$
transport, and an outer porous organic layer penetrable by solvent molecules,
have been demonstrated.\cite{harris,peled,andersson}  Note that as-formed SEI
films may further undergo chemical/electrochemical evolution during cycling.

To model two-$e^-$ reactions, we use Li$_{13}$Si$_4$ model anodes with
their (010) surfaces in direct contact with a liquid mixture of FEC and EC.
This surface is meant as a prototype to study the fast, two-$e^-$ electron
transfer regime.  It is meant to mimic battery charging-induced Si cracking
events that expose SEI-free surfaces to the electrolyte.  This can take 
place during the initial or subsequent charge cycles.
The anode stochiometry is chosen to represent one of the strongly
lithiated Li-Si crystalline phases,\cite{dahn} which are in turn models for
disordered Li-Si alloys in cycled materials.  Unlike the more commonly studied
Li$_{15}$Si$_4$ stoichiometry, the orthonormal unit cell of crystalline
Li$_{13}$Si$_4$ exhibits three unequal lattice constants.  This inequality
facilitates lattice-matching to model oxide films on anode surfaces.
Rapid FEC decomposition induced by two $e^-$ transfer is observed.
One-electron decomposition of FEC is examined using cluster-plus-dielectric
models in the absence of the silicon electrode.  Such calculations were
pioneered by Balbuena and coworkers and are valid for reactions that take
place in the bulk liquid region.\cite{bal01,bal02,bal02a} 

We also address the role played by silicon oxide.  Theoretical studies of
silica and related oxides in the context of lithium-ion batteries have been
conducted,\cite{ban,garo,vanduin} but SEI film formation reactions on those
surfaces were not a focus.  It may be argued that silicon anodes
{\it should} qualitatively differ from graphite with respect to initial
SEI formation.  Unlike graphite, silicon samples are generally covered with
a native oxide layer about 20~\AA\ thick.  Recent work demonstrated that
SEI films on cycled silicon anodes contains buried SiO$_2$ and/or lithium
silicate derived from SiO$_2$.\cite{edstrom}  This dense oxide/silicate
insulating layer may slow down $e^-$ transfer sufficiently during initial
charging that the two-electron regime entirely vanishes.\cite{insulating}
Therefore a computational model that includes an electrode as well as a
passivation oxide layer is also considered.

This paper is organized as follows.  Section~2 briefly describes the
computational methods.  FEC decompositions induced by two- and
one-electron injections are described in Sec.~3 and Sec.~4, respectively.
Sec.~5 discusses the implications of our findings for SEI compositions and
provides comparisons with experiments, and Sec.~6 concludes the paper with
a brief summary.  Two appendices examine the reductive decomposition of
FEC adsorbed on lithium and silicon clusters using hybrid DFT functionals,
and report two-electron mechanisms similar to those described in the main text.

\section{Methods}
\label{method}

Finite temperature {\it ab initio} molecular dynamics
(AIMD) simulations are conducted under solvent-immersed
electrode conditions to study two-electron-induced FEC decomposition.
These calculations are performed using the Vienna Atomic
Simulation Package (VASP) version 4.6\cite{vasp,vasp1} and the PBE density
functional.\cite{pbe}  A 400~eV planewave energy cutoff and a
10$^{-6}$~eV convergence criterion is applied at each Born-Oppenheimer
time step.  Tritium masses on EC are substituted for protons
to permit a time step of 1~fs.  The trajectories are kept at an average
temperature of T=450~K using Nose thermostats.  The elevated temperature
reflects the need to ``melt'' EC, which has an experimental freezing point
above room temperature, and to improve sampling efficiency.\cite{water}  In
real batteries DMC cosolvent molecules reduce the viscosity, but DMC is
not included herein.  As a result, bond-breaking reactions observed
in this work reflects accelerated kinetics pertinent to T=450~K.

Several types of simulation cells are considered.  One has dimensions
15.84$\times$33.00$\times$13.32\AA$^3$ and contains an apolar
Li$_{156}$Si$_{48}$ slab with a zero dipole moment perpendicular to the
exposed (010) surfaces.  The bulk Li$_{13}$Si$_4$ structure is taken from
Ref.~\onlinecite{dahn}.  For that system, the (010) surface is found to be
among the lowest energy low index surfaces.\cite{martinez} The vacuum
gap between the exposed surfaces is filled with 32~EC molecules, which are
pre-equilibrated using Monte Carlo simulations with the Towhee Monte Carlo code,
and simple molecular force fields described in earlier works.\cite{pccp,ald}
Then 8 of the 32 EC molecules at the electrode surfaces are replaced with FEC
and Monte Carlo is resumed for 5000 passes.  This spatial distribution of
molecules faciliates FEC decomposition over EC.  Finally, AIMD is initiated
using 2$\times$1$\times$2-point Brillouin zone sampling.
During experimental preparation of battery cells, the voltage is lowered
slowly. The higher reductive potentials of additives mean that they
are first decomposed before other electrolyte components are reduced.
(See however Ref.~\onlinecite{okuno} for an alternate interpretation.)  Strict
voltage control is lacking in these AIMD simulations,\cite{voltage} which 
are meant to mimic post-Li$_x$Si$_y$ cracking during cycling conditions, not
electrode preparation.  Thus, we have used spatial variations in additive
compositions to enhance preferential FEC decomposition.  In VASP calculations,
the Bader decomposition technique is applied to determine the net charge on
FEC fragments.\cite{bader}

Another simulation cell has the same Li-Si alloy anode and liquid electrolyte,
but includes a $\sim 7$~\AA\, thin, stoichiometric Li$_4$SiO$_4$ layer
on each electrode surface.  Each oxide film amounts to a bilayer of
SiO$_4^{4-}$ units surrounded by Li$^+$ ions.  The cell dimensions
are 15.84$\times$44.40$\times$13.32\AA$^3$.  Details about 
this oxide-covered model anode will be discussed in future publications.

Finally, we consider an AIMD trajectory without an electrode.  31~EC, 1~FEC,
1~Li$^+$, and an excess electron reside in this (15.24\AA)$^3$ simulation
cell.  A liquid cell configuration of 32~EC and Li$^+$ is taken from
Ref.~\onlinecite{pccp}.  One EC coordinated to Li$^+$ is replaced by an FEC
molecule which is deformed into a geometry that resembles an FEC$^-$ (see the
Results sections below), and then an $e^-$ is added.  When AIMD is initiated
from this configuration, the excess $e^-$ is found to be successfully
localized on the FEC molecule.

Static cluster-based, static optimization calculations apply DFT/PBE and
M\"{o}ller-Plesset second order perturbation (MP2) in conjunction with the
Gaussian (G09) suite of programs\cite{g09} and the ``smd'' dielectric continuum
approximation.\cite{smd}  The dielectric constant $\epsilon$ is set to 40,
which approximates the environment of the entire battery electrolyte,
not just liquid EC.  Geometry optimization is performed using the
{\tt 6-31+G(d,p)} basis set.  Single point energies are computed at the
{\tt 6-311++G(3df,2pd)} level of theory.  Vibrational
frequencies are calculated for stable structures and transition states using
the smaller basis, yielding zero point energies and thermal corrections to
electronic energies.  The DFT/PBE functional is chosen to complement AIMD/PBE
simulations in this work, and the MP2 method is picked to be consistent with
MP2 calculations in one of the authors' previous work\cite{e2} and other 
modeling effort in the literature.\cite{han04,bedrov}  Both
(EC)$_2$FEC$^-$:Li$^+$ and FEC$^-$:Li$^+$ clusters are considered.  The larger
cluster is chosen to reflect a
previous AIMD prediction that Li$^+$ is coordinated to slightly more than 3~EC
molecules in EC liquid if one  EC contains an excess electron.\cite{pccp}

Hybrid DFT functionals\cite{becke,lyp,pw91} are used for studies of
FEC:Li$_{50}$ and FEC:Li$^+$:Si$_{16}$H$_{15}$ clusters.  The results are
described in the appendices.

\section{Results: Two-Electron Mechanism}

\subsection{Pristine Electrodes}
\label{twoelectron}

\begin{figure}
\centerline{\hbox{ \includegraphics*[width=2.35in]{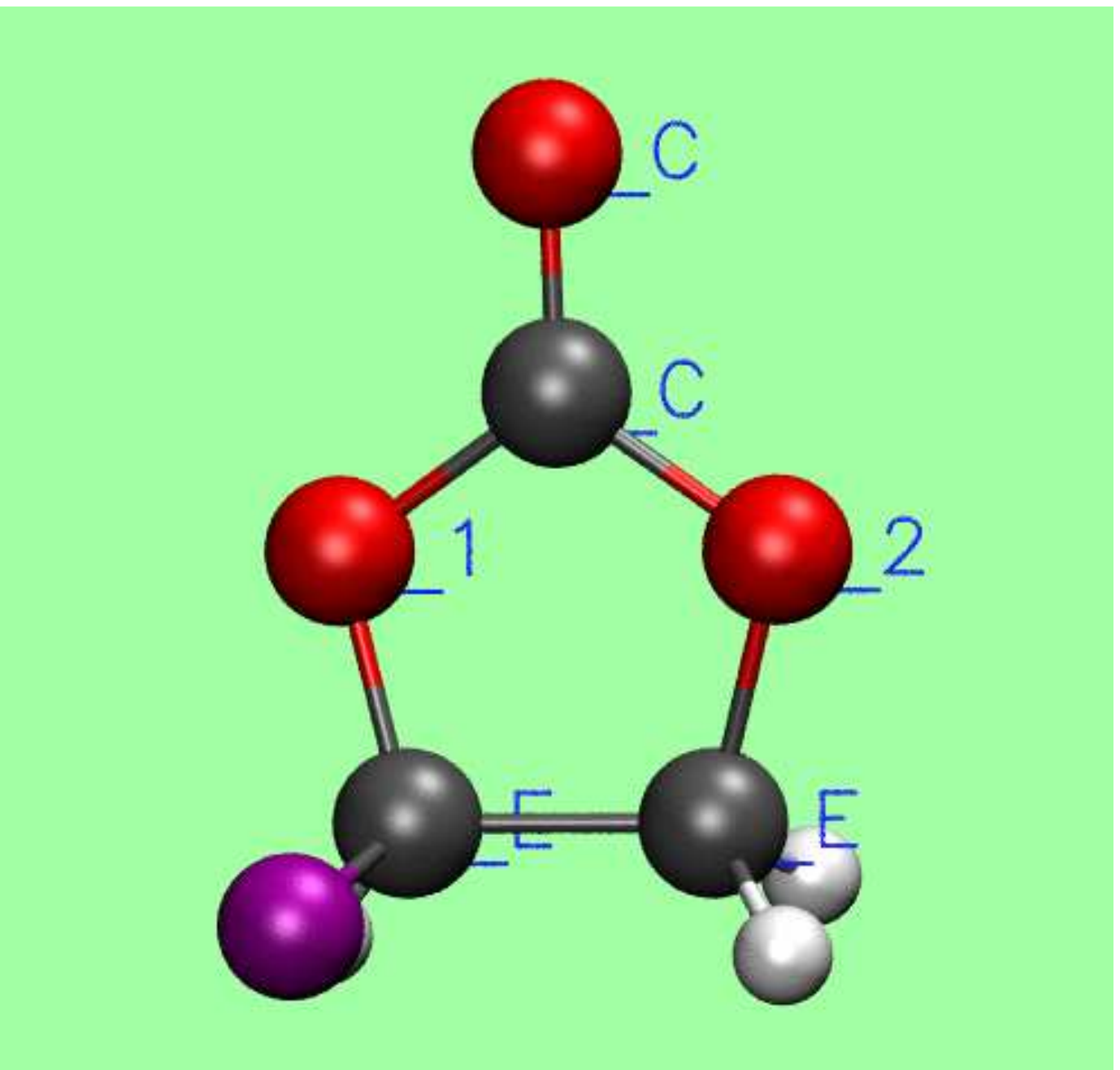}
		   \includegraphics*[width=2.15in]{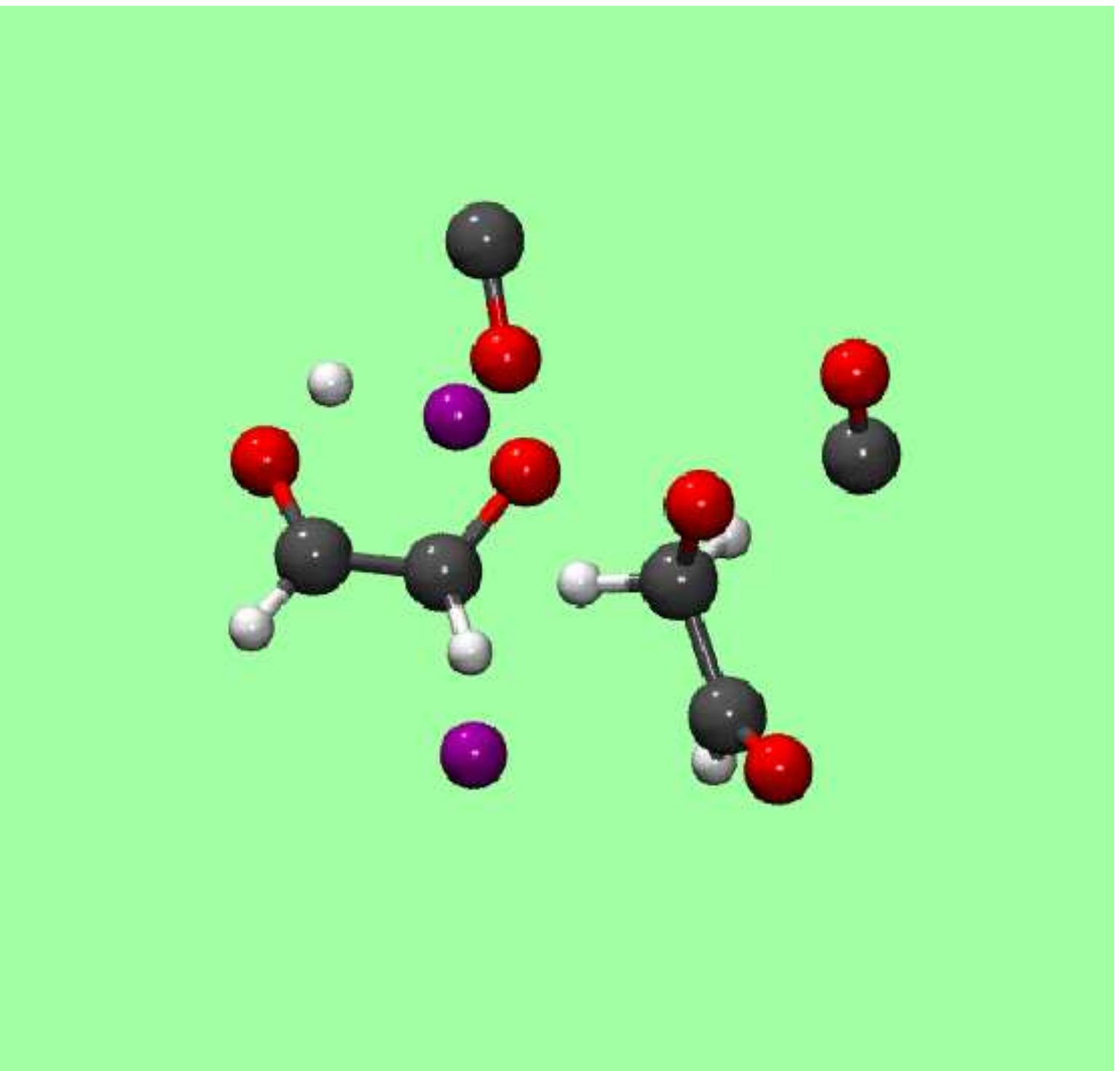}
		   \includegraphics*[width=2.15in]{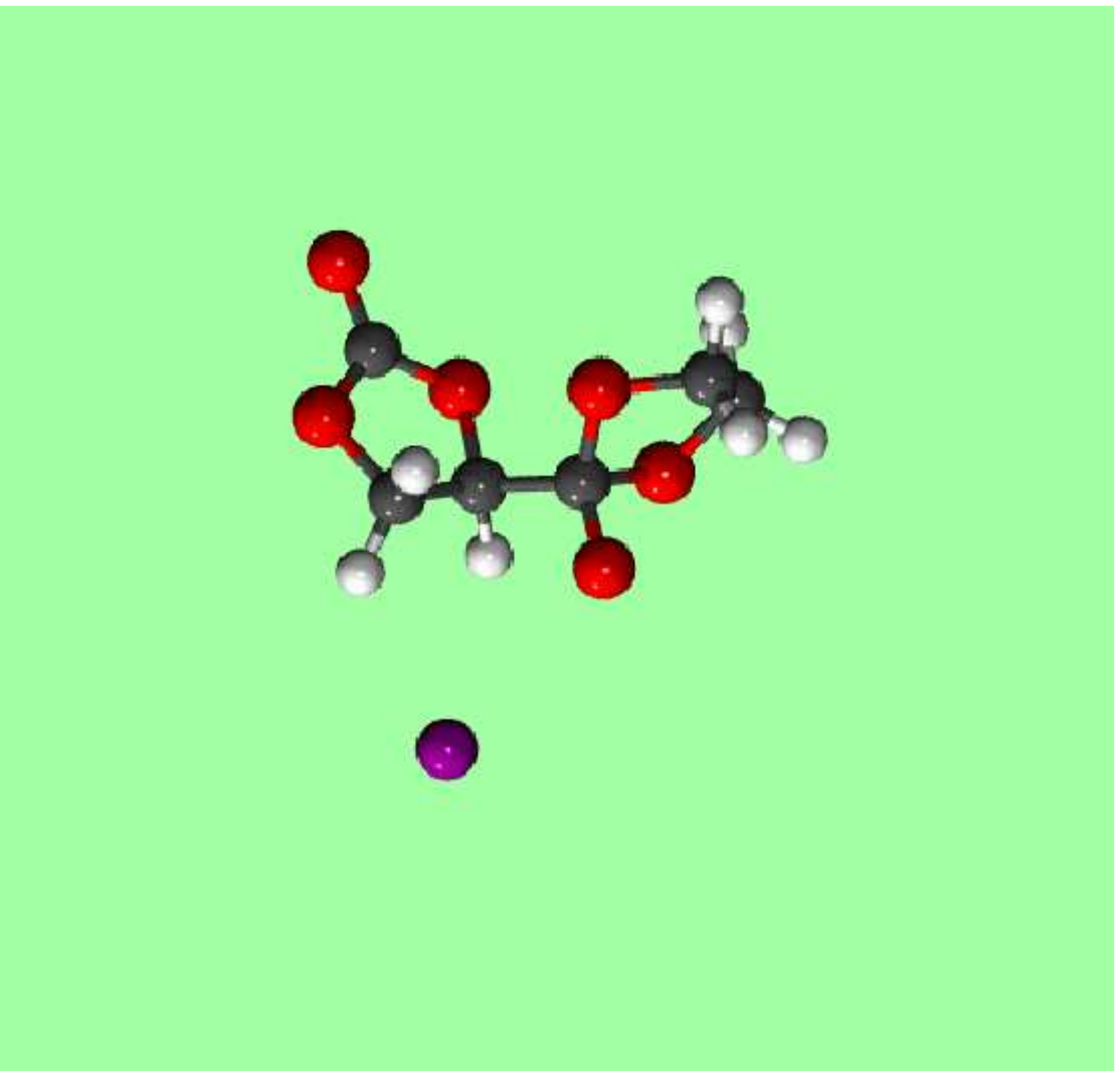} }}
\centerline{\hbox{ (a) \hspace*{1.95in} (b) \hspace*{1.95in} (c)}}
\centerline{\hbox{ \includegraphics*[width=2.55in]{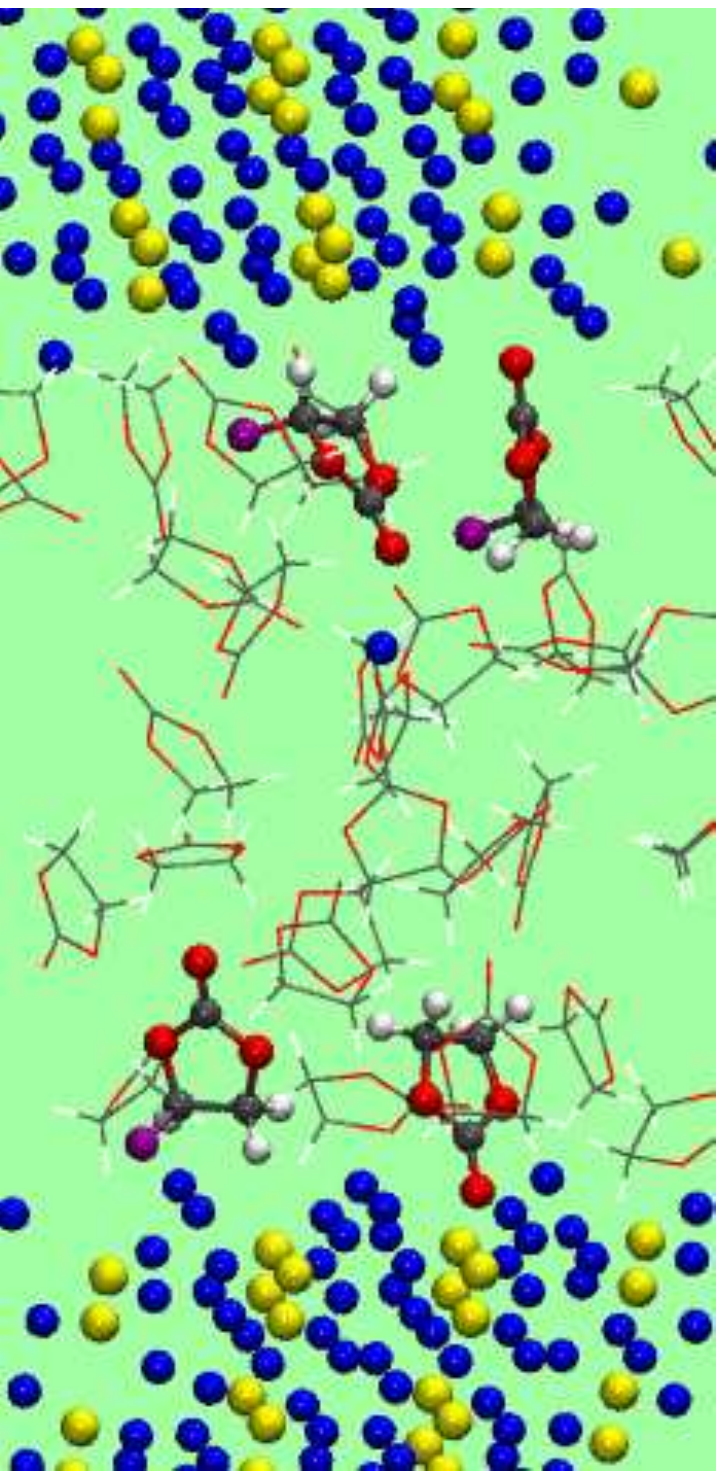}
		   \includegraphics*[width=2.55in]{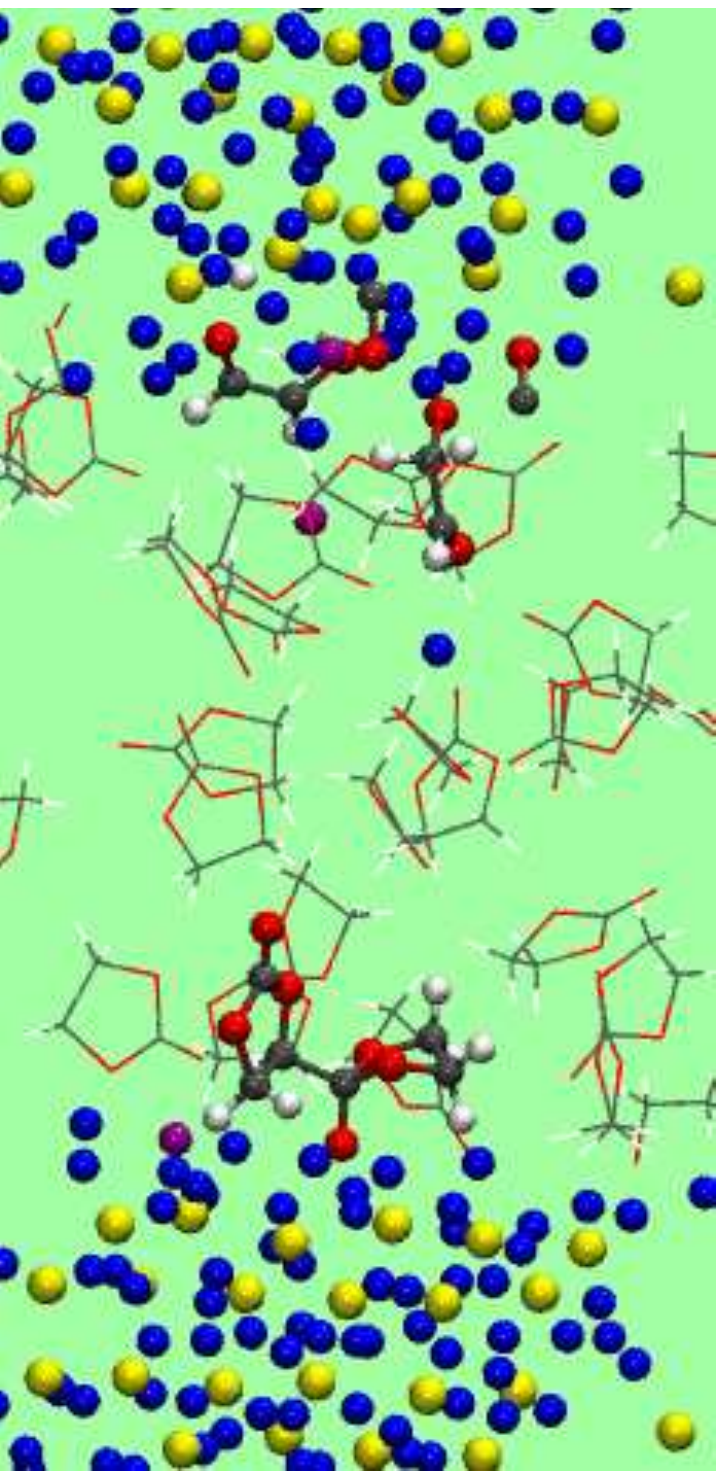}
		   \includegraphics*[width=2.55in]{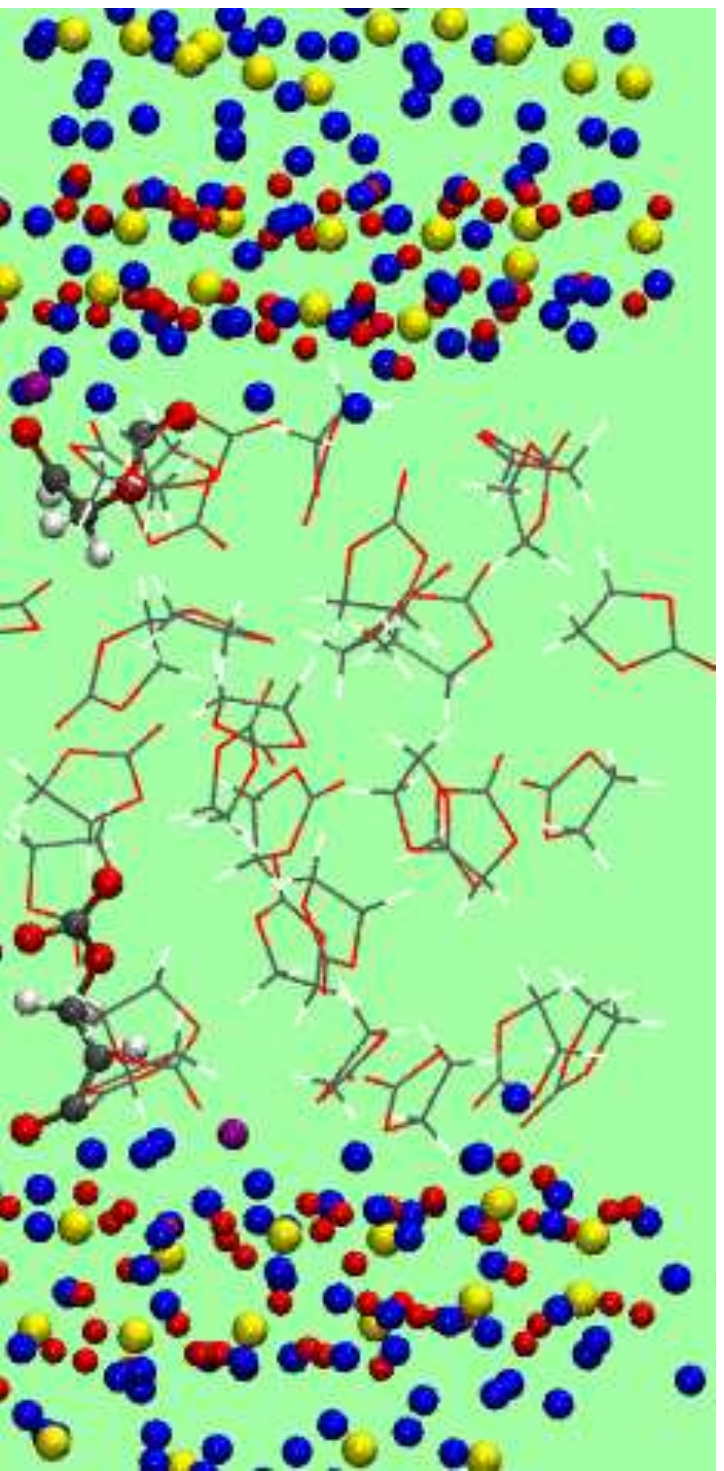} }}
\centerline{\hbox{ (d) \hspace*{1.95in} (e) \hspace*{1.95in} (f)}}
\caption[]
{\label{fig1} \noindent
(a) FEC molecule with labels.  C, O, H, and F atoms are depicted as grey,
red, white, and purple spheres.  (b)\&(c) Expanded views of decomposed
molecules on the top and bottom sides of panel (e), omitting the
Li$_{13}$Si$_4$ slab.  (d)\&(e) Configurations at times $t=0$ and $t=4$~ps
into an AIMD simulation of a Li$_{\rm 13}$Si$_4$ slab immersed in liquid
FEC/EC.  Si, Li, C, O, H, and F atoms are depicted as yellow, blue, grey,
red, white, and purple lines or spheres.  In (d), 3~FEC and 1~EC molecules at
the surface, about to react, are depicted as ball-and-stick models.  In (e),
these 3~FEC and 1~EC that have reacted.  Other reactions are omitted.
(f) Liquid FEC in contact with Li$_4$SiO$_4$ coated Li$_{13}$Si$_4$
anode slab, also leading to two-$e^-$ reduction reactions.
}
\end{figure}

We first consider unbiased AIMD simulations of liquid electrolyte in contact
with the (010) surfaces of a Li$_{13}$Si$_4$ slab.  Figs.~\ref{fig1}d
and~\ref{fig1}e depict snapshots at the $t=0$ and $t=4$~ps points of an
AIMD trajectory.  On the upper surface of Fig.~\ref{fig1}e, three FEC
molecules have spontaneously absorbed two $e^-$ from the electrode, each
breaking first one C$_{\rm C}$-O$_1$ or C$_{\rm C}$-O$_2$ bond and then the
other (two are shown in Fig.~\ref{fig1}e).  An F$^-$ anion is ejected almost
simultaneously.  This leaves CO, F$^-$, and C$_2$H$_3$O$_2^-$ as products.
Fig.~\ref{fig1}b zooms in on these fragments.  In one of them, a C-H bond
has also cleaved, releasing a hydrogen atom that becomes a H$^-$ bound to
a Si atom on the anode surface (Fig.~\ref{fig1}b).  Even in this case, the
product cannot be considered ``HF.'' On the bottom surface (Fig.~\ref{fig1}e),
a F$^-$ is ejected before any other bond
breaking occurs.  The resulting C$_3$H$_3$O$_3^-$ anion then
initiates nucleophilic attack on the C$_{\rm C}$ atom of a neighboring
intact EC molecule (see Fig.~\ref{fig1}c for details).  This may be the
initiation of oligomer formation.  No HF or CH$_2$CHF is released, contrary
to mechanisms proposed in the literature.\cite{japan,vc} By the nature of
the simulation, the decomposition reactions observed are not imposed by
any assumption, but are the natural manifestations of low-barrier mechanisms.

Along this trajectory, after another 5.6~ps, two EC and another FEC molecule
on the surfaces decompose (not shown).  The two EC molecules absorb two
$e^-$ each to form CO/C$_2$H$_4$O$_2^{2-}$ and C$_2$H$_4$/CO$_3^{2-}$, as
has been observed on Li$^+$ intercalated graphite edges.\cite{pccp} The
newly reacting FEC forms CO, C$_2$H$_3$O$_2^-$, and F$^-$, the majority
product in Fig.~\ref{fig1}e.  No further dimerized product is observed.
While we have only considered one AIMD trajectory with a single initial,
force-field pre-equilibrated configuration, the multiple bond-breaking
events observed, to some extent, serve as a reasonable sized sample of
chemical reactions.  The majority of FEC molecules, upon accepting two $e^-$,
react to form CO and F$^-$.

On the whole, these FEC~+~2~$e^-$ reactions yield more varied
products/intermediates than EC$^{2-}$ breakdown.\cite{e2} Simulations of FEC
reduction on a lithium cluster and cluster models of weakly-lithiated Si
surfaces (appendices A and B) show that C$_{\rm C}$-O bond breakings via
2 $e^-$ mechanisms are also thermodynamically and kinetically favorable,
facilitated by multiple bindings with the surface.  Other products, including
Si-C bonded motifs, are observed if a different electrode stochiometry
is used in AIMD simulations (see Ref.~\onlinecite{martinez} for details).

Among the fragments observed on these Li$_{13}$Si$_4$ surfaces, F$^-$
should precipitate with Li$^+$ to form LiF(s), which is a part of the
SEI.  The R-CHO$^-$ functional groups of the organic anionic fragments
(Fig.~\ref{fig1}b) are reactive and may participate in further nucleophilic
attacks on intact solvent molecules, like the 2-$e^-$ decomposition products
of EC.\cite{e2}  Unlike 2-$e^-$ attack on EC,\cite{bal01,e2} CO$_3^{2-}$
formation is not observed.  Appendix~B also shows that a 2-$e^-$ induced
CO$_3^{2-}$-releasing FEC reaction is exothermic but exhibits a high barrier
when computed on a Si-cluster.  Unlike EC$^{2-}$,\cite{e2} we have been unable
to stabilize intact FEC$^{2-}$ molecular structures in a cluster geometry.
FEC$^{2-}$:Li$^+$ complexes always spontaneously decomposes, whether a
dielectric continuum treatment is used or not.  Hence no cluster-based
calculation of FEC$^{2-}$ bond-breaking barrier/reduction potential or
direct comparison with AIMD/PBE reaction timescales following 2-$e^-$
reduction of intact FEC molecules can be made.

\subsection{Attempt at Modeling One-$\it e^-$ Mechanisms on
Oxide-Covered Electrodes}

We have made a preliminary attempt at modeling 1-$e^-$ attacks on FEC
in the presence of a Li$_{\rm 13}$Si$_4$ electrode, with each surface
coated with a 7~\AA\, thick Li$_4$SiO$_4$ layer so that the FEC liquid
is no longer in contact with Li$_{13}$Si$_4$ surfaces (Fig.~\ref{fig1}f).
Two FEC are decompose in picosecond time scales.  Despite the
expectation that the oxide should slow down electron transfer and enhance
1-$e^-$ attack on FEC over 2-$e^-$ reduction, Bader charge decomposition
analysis shows that the products are still consistent with 2-$e^-$-induced
reactions.  

The reason for 2-$e^-$ attack may be twofold.  (1) The reduction potentials
may favor double reduction of FEC, as in the case for EC.\cite{e2}  The
proximity of the metallic electrode means that the dielectric constant at high
frequencies remains large.  High $\epsilon_\infty$ means that reorganization
energies for both one and two $e^-$ transfer to FEC may be low, and therefore
electron transfer through the oxide is fast.  If that is indeed the case, two
electron attack is the correct prediction.  (2) However, AIMD/PBE
overestimates $e^-$ tunneling rate due to DFT/PBE delocalization errors.
This tendency has been discussed in Ref.~\onlinecite{ald}.  Thus,
2-$e^-$ attack on oxide-covered anode may be an incorrect DFT prediction.
Hybrid DFT functionals, less susceptible to DFT delocalization errors, may
be used to study this system in the future, although they
are costly to use in the periodically replicated simulation cells discussed
in this section.  In the next sections, 1-$e^-$ mechanisms are imposed
by adding one excess $e^-$ to model systems in the absence of an electrode.

\section{Results: One-electron Mechanisms}

\subsection{AIMD simulations in bulk liquid}
\label{aimd1e}

An AIMD trajectory of 31~EC, 1~FEC, 1~Li$^+$, and an excess $e^-$ in the
simulation cell is considered next.  By design, the excess $e^-$ resides on
the FEC molecule which is coordinated to the Li$^+$, as discussed in
Sec.~\ref{method}.  After 5.3~ps, the initially intact FEC$^-$ breaks a
C$_{\rm C}$-O$_1$ bond, similar to the initial reaction step observed in
most reactions on Li$_{\rm 13}$Si$_4$ surfaces (Fig.~\ref{fig1}e).  
A snapshot of this configuration is shown in Fig.~\ref{fig2}.  
Using a different force-field pre-equilibrated initial configuration, the
same bond breaks after 16.0~ps (not shown).  The bond-breaking time scales are
slightly different due to the stochastic nature of barrier crossing events.
Intuitively, C$_{\rm C}$-O$_1$ bond-breaking should be favored over
C$_{\rm C}$-O$_2$ because the electron-withdrawing F-atom should stabilize
the -CHFO$_1^-$ group.  Unlike two-electron-induced
reactions described in the previous section, the C-F bond has not 
broken and the F$^-$ has not yet detached from the FEC$^-$ in this
trajectory.  The reason for the intact C-F bond may be the lack of Li$^+$ in
the proximity to abstract F$^-$ and form a stable LiF cluster.  This point
will be reiterated in light of cluster-based calculations described below.  

\subsection{Cluster Calculations: Intact FEC$^-$}
\label{cluster}

Cluster-based predictions are depicted in Figs.~\ref{fig3}-\ref{fig6}.  First
we consider the geometries of metastable FEC$^-$:Li$^+$ complexes prior to
bond-breaking events.  Configurations {\bf A}-{\bf D} (Fig.~\ref{fig3}a-d) are
somewhat analogous to the EC$^-$:Li$^+$ geometry.\cite{bal01}  The
C$_{\rm C}$-O$_{\rm C}$ bond is bent out of the plane containing the 5-atom
ring, like EC$^-$, but unlike charge neutral FEC or EC, because the
C$_{\rm C}$ atom is now sp$^3$ (or ``sp$^{2.5}$'') hybridized
to accommodate an extra $e^-$.  Unlike the more symmetric EC$^-$:Li$^+$,
complex, there are four distinct binding geometries depending on whether
Li$^+$ and F~reside in the same or opposite side of the 5-member ring, and
whether they are on the same side of the bisecting plane perpendicular to the
ring.  Another configuration, {\bf E} (Fig.~\ref{fig3}e), finds Li$^+$ residing
on the approximate bisector plane, only coordinated to O$_{\rm C}$.  All~5
configurations are close in energy, within 0.065~eV (0.042~eV) of each other
depending on whether MP2 (PBE) is used (Table~\ref{table1}).  Configuration
{\bf A}~(Fig.~\ref{fig3}a) is most stable according to the PBE functional.
When MP2 is used, it is more stable than all other complexes except {\bf D}
(Fig.~\ref{fig3}d) by a small 0.020~eV, which is within thermal energy of
the MP2-predicted global minimum.

Given the fact that neither DFT nor MP2 has attained chemical accuracy
(i.e., a mean error of less than $\sim$0.04~eV compared to experiments),
we will henceforth use {\bf A}~as the starting configuration of all further
cluster-based degradation studies.  The most important metrics are the
barrier heights measured against the most stable intact FEC$^-$ configuration.
A small uncertainty of $\sim$0.02~eV in the energy of the starting, unreacted
complex merely adds a small amount to barrier heights and does not change
the qualitative conclusions.

Using Configuration {\bf A} to represent FEC$^-$:Li$^+$, the predicted MP2
reduction potential of FEC:Li$^+$ is 0.75~V, which is 0.22~V higher than
that of EC:Li$^+$ based on a previous study at the same level of
theory.\cite{e2} This difference indicates that FEC:Li$^+$ accepts an 
electron at higher voltages than EC:Li$^+$.  The experimental value appears
to be $\sim$ 0.95~V.\cite{fec_redox} As mentioned above, adding a second
$e^-$ to the FEC$^-$:Li$^+$ cluster leads to immediate decomposition, and 
thus the reduction potential of intact FEC$^-$ cannot be computed.

If the dielectric continuum approximation is omitted, the C$_{\rm C}$-O$_1$
bond in {\bf A} spontaneously breaks during geometry optimization regardless
of whether PBE or MP2 is used.  The resulting structure is similar to {\bf F}
(Fig.~\ref{fig3}f, which is however optimized with a dielectric approximation,
not in the gas phase).  Recall that the same bond spontaneously breaks in AIMD
simulations (Fig.~\ref{fig2}).  The same bond also breaks in {\bf C} and
{\bf D} in the gas phase (no dielectric) during PBE optimization, while the
C$_{\rm C}$-O$_2$ bond does not.  The next section focuses on decomposition
initiated by C$_{\rm C}$-O$_1$ bond-breaking.

\subsection{Cluster Calculations: Proposed FEC$^-$ Decomposition Mechanisms}

This subsection shows that the most probable (one-eletron) FEC$^-$
decomposition route is an indirect, multi-step reaction leading to removal
of an F$^-$ anion followed by release of CO$_2$ and an organic radical.

\begin{figure}
\centerline{\hbox{ \includegraphics*[width=3.in]{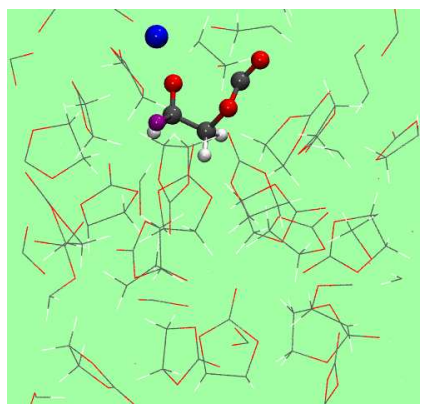} }}
\caption[]
{\label{fig2} \noindent
Configuration at the end of a 5.3~ps AIMD trajectory in the absence of
an electrode.  Solid spheres highlight the breaking of the C$_{\rm C}$-O$_1$
bond.  Li$^+$ is coordinated to between 3 and 4 O-atoms on
FEC$^-$ and EC molecules (the latter shown as lines).
}
\end{figure}

\begin{figure}
\centerline{\hbox{ \includegraphics*[width=2.15in]{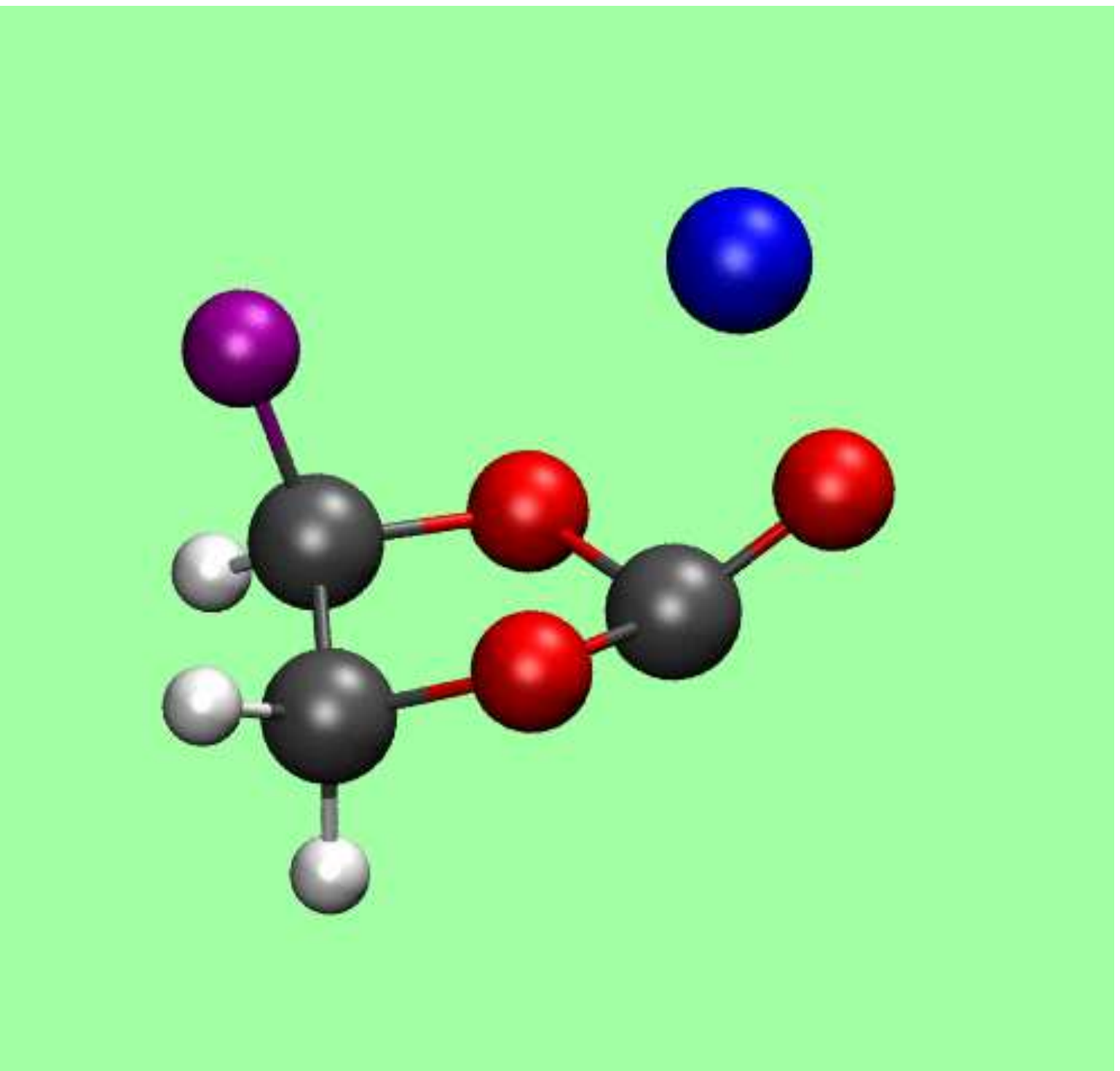}
		   \includegraphics*[width=2.15in]{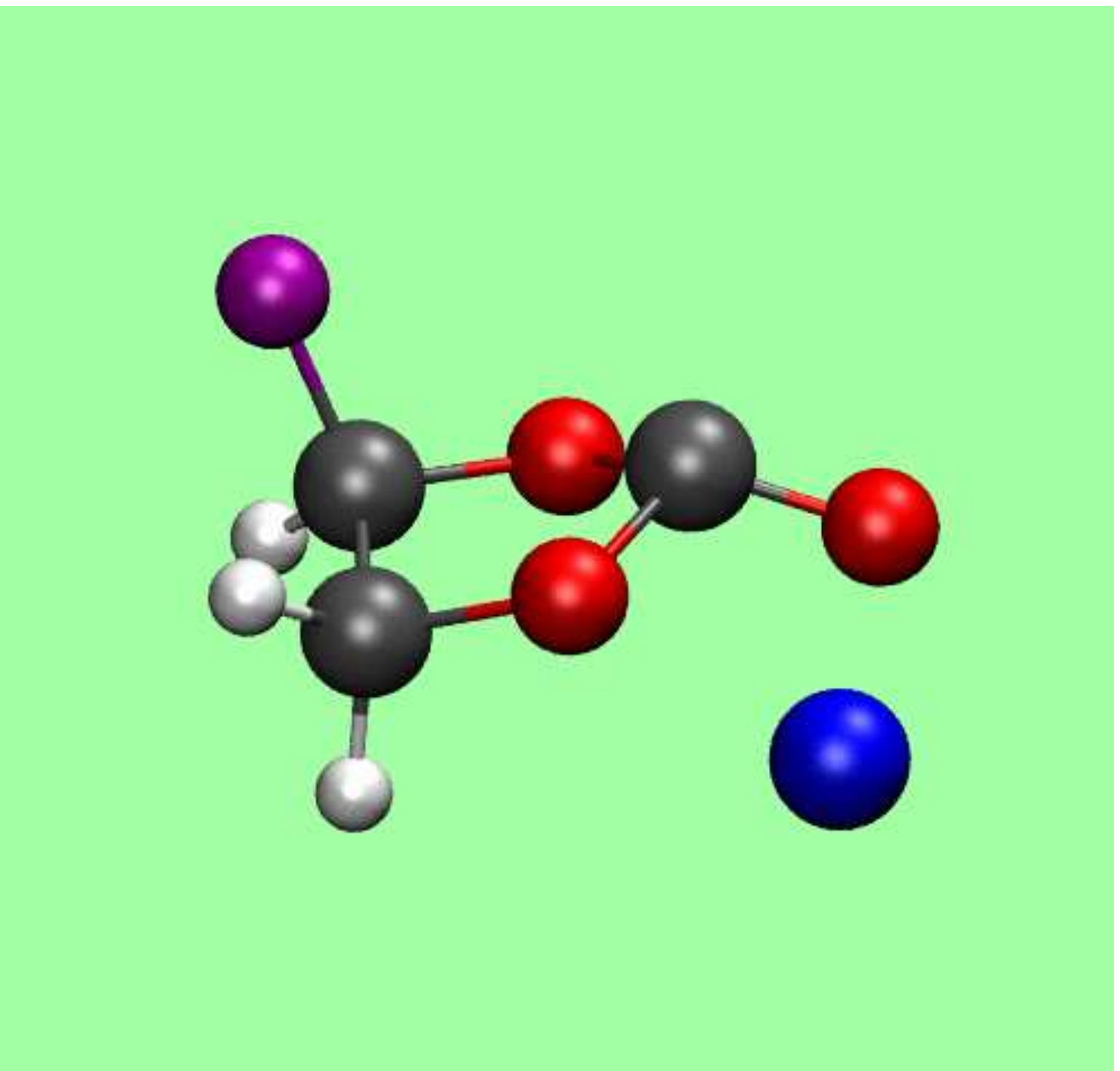}
		   \includegraphics*[width=2.15in]{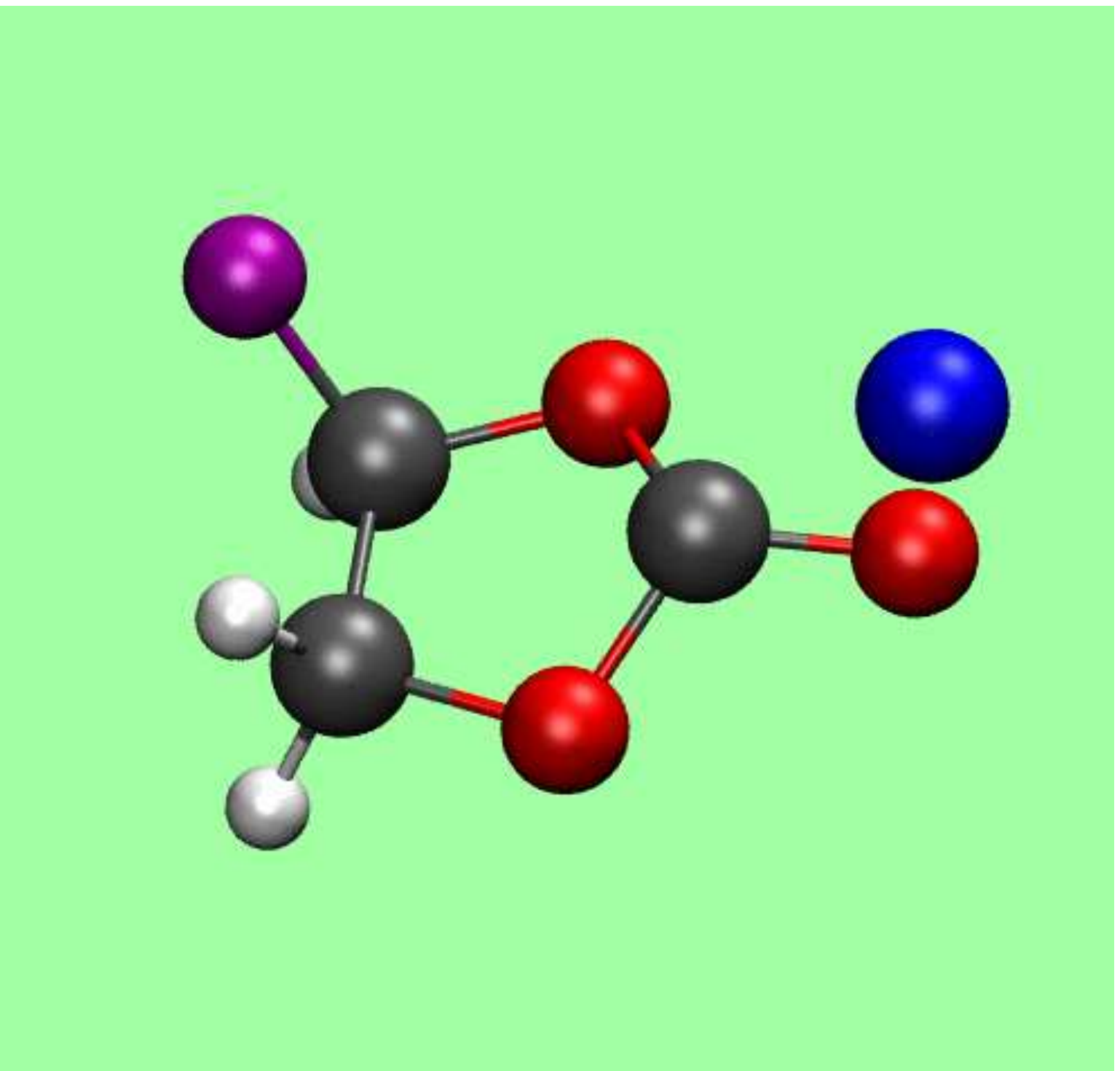} }}
\centerline{\hbox{ (a):A \hspace*{1.85in} (b):B \hspace*{1.85in} (c):C}}
\centerline{\hbox{ \includegraphics*[width=2.15in]{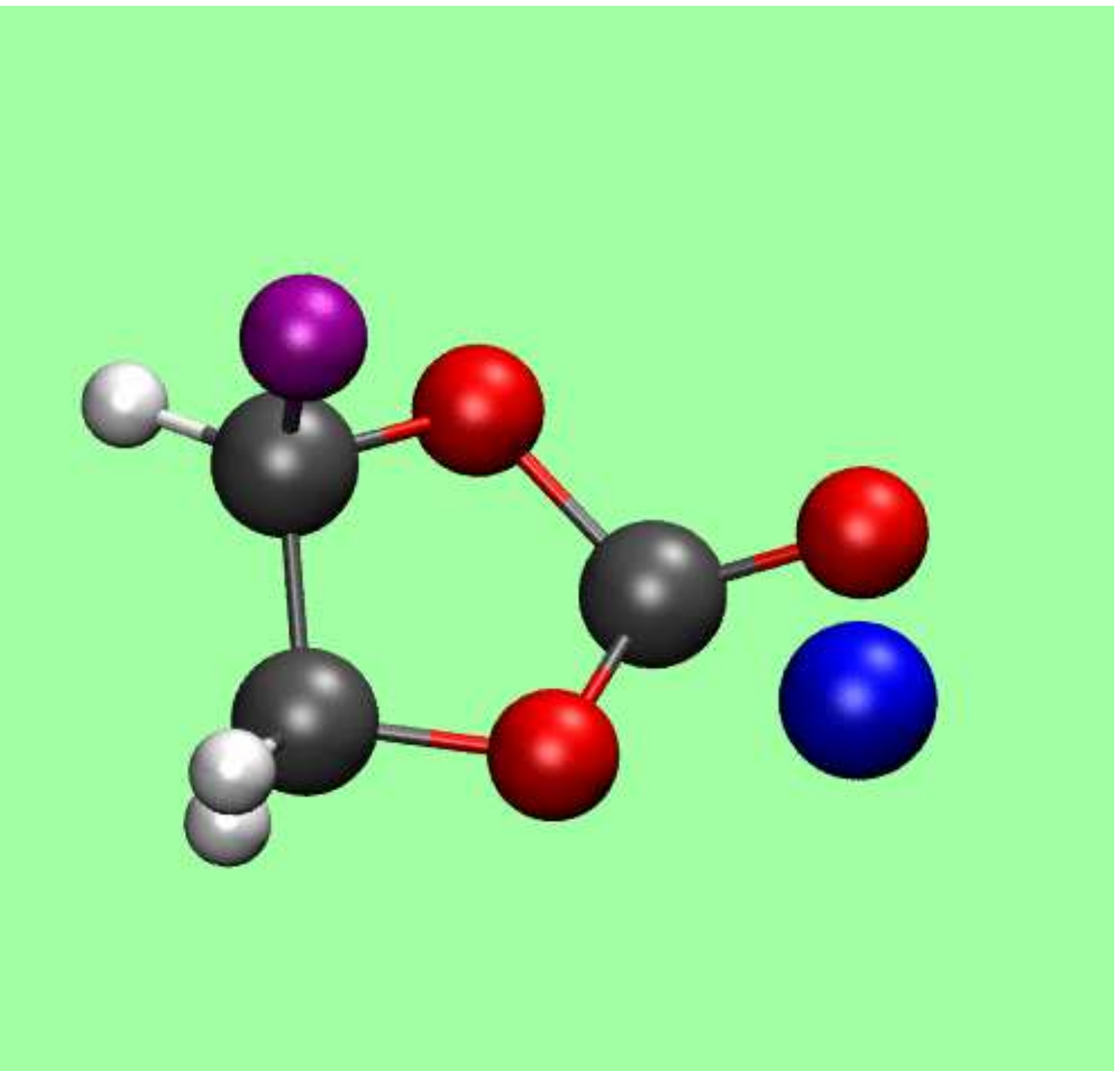}
		   \includegraphics*[width=2.15in]{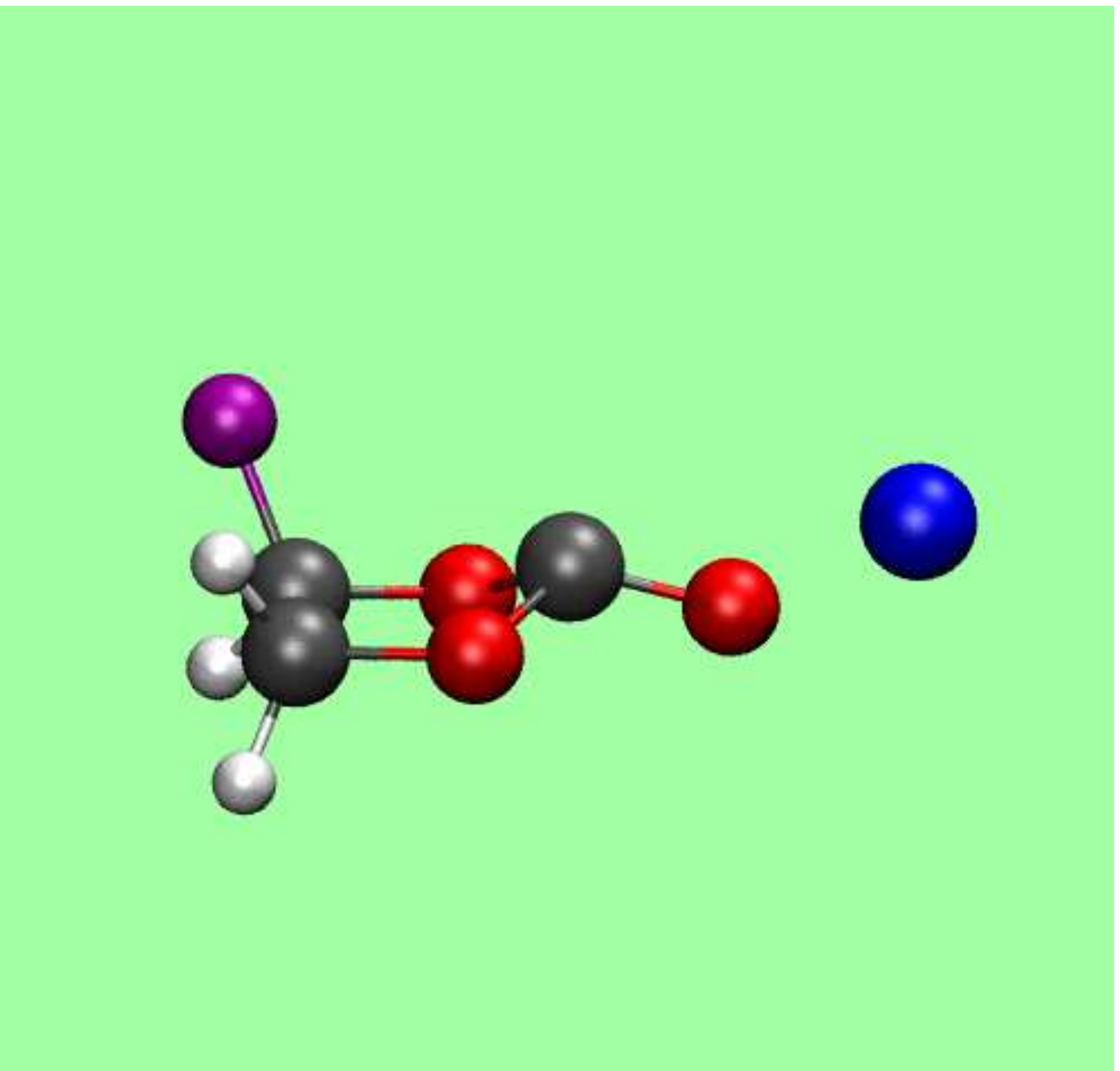}
		   \includegraphics*[width=2.15in]{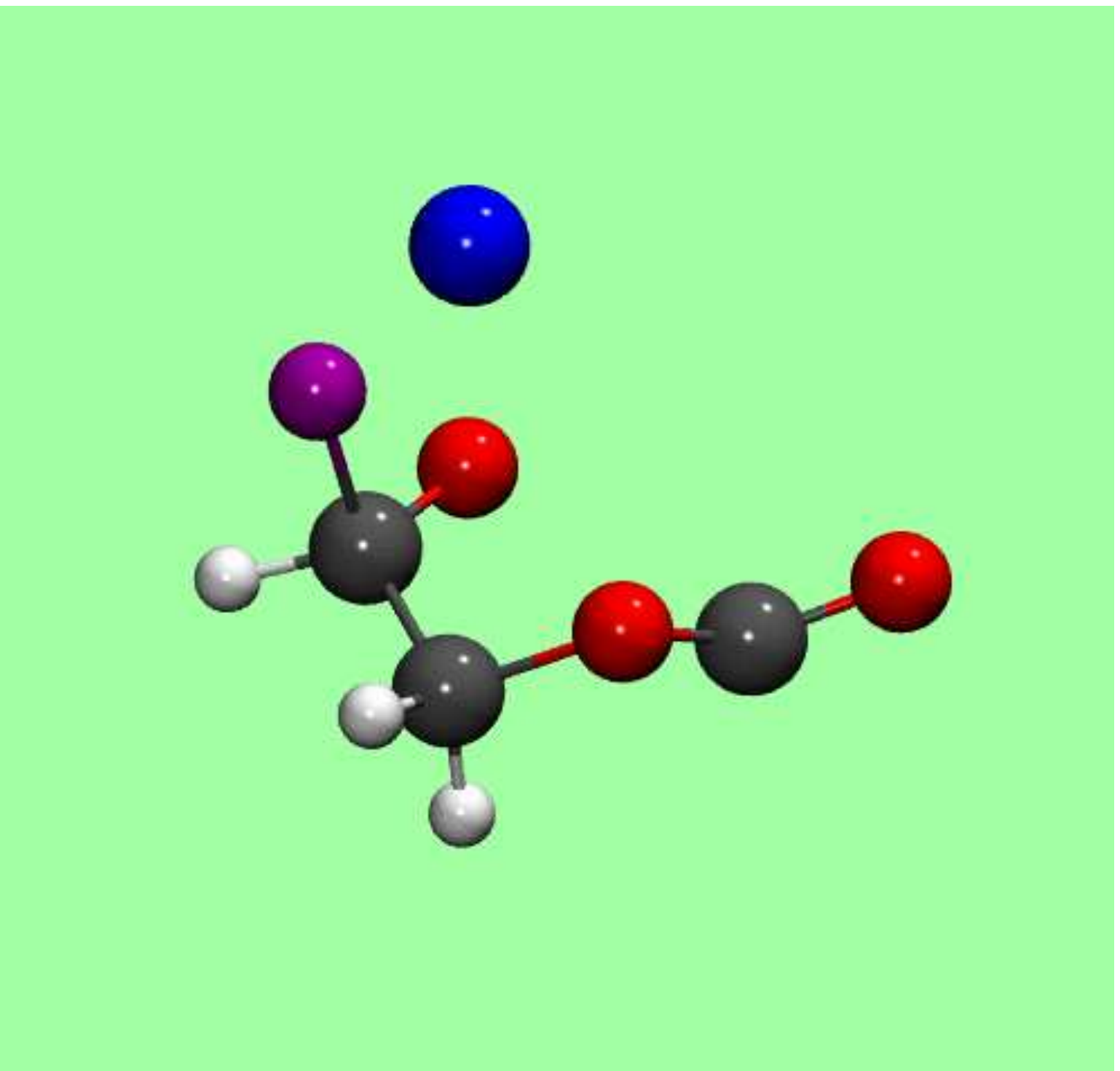} }}
\centerline{\hbox{ (d):D \hspace*{1.85in} (e):E \hspace*{1.85in} (f):F}}
\caption[]
{\label{fig3} \noindent
Panels (a)-(f): FEC$^-$:Li$^+$ clusters.  Configurations {\bf A}-{\bf F}.
Li$^+$ is coordinated to two O atoms of FEC$^-$ in panels (a)-(d) and
only one O~in panel (e).
}
\end{figure}

Bond-breaking barriers computed using the FEC$^-$:Li$^+$ cluster are listed
in Table~\ref{table2}, and the global free energy landscape is illustrated
in Fig.~\ref{fig5}.  C$_{\rm C}$-O$_{\rm 1}$ cleavage yields the lowest
initial bond-breaking barrier, predicted using  the MP2 (PBE) method
to be 0.26~eV (0.23~eV).  This forms intermediate {\bf F} (Fig.~\ref{fig3}f)
which is metastable by 0.09~eV (stable by $-0.11$~eV).  The small
($\sim 0.1$~eV) discrepancies between MP2 and PBE barriers indicate that these
reactions should be fast regardless of computational methods.  A fast
reaction rate is qualitatively consistent with the AIMD trajectory
(Fig.~\ref{fig2}) showing C$_{\rm C}$-O$_1$ bond-breaking within picoseconds.  

\begin{figure}
\centerline{\hbox{ \includegraphics*[width=2.15in]{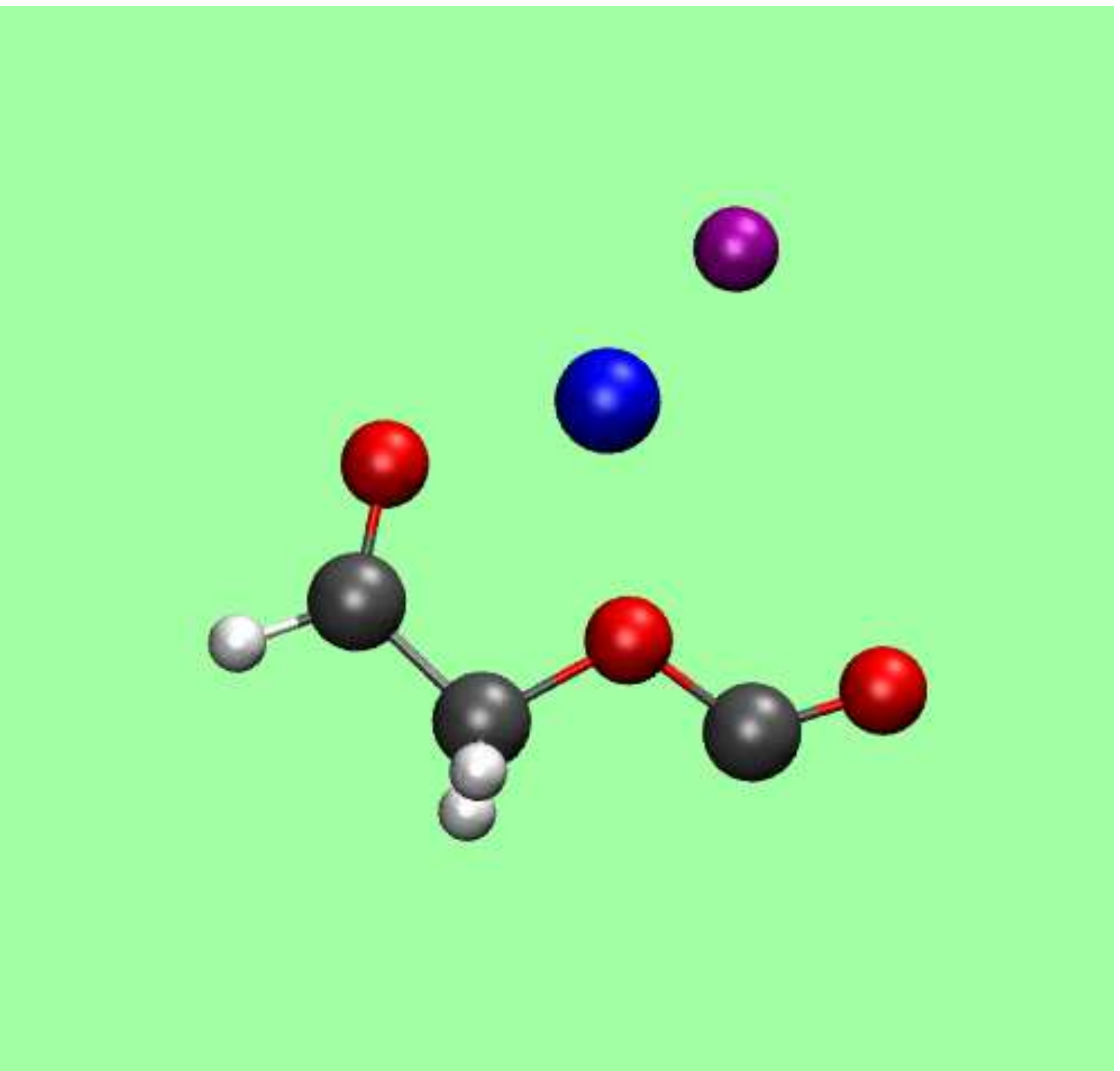}
		   \includegraphics*[width=2.15in]{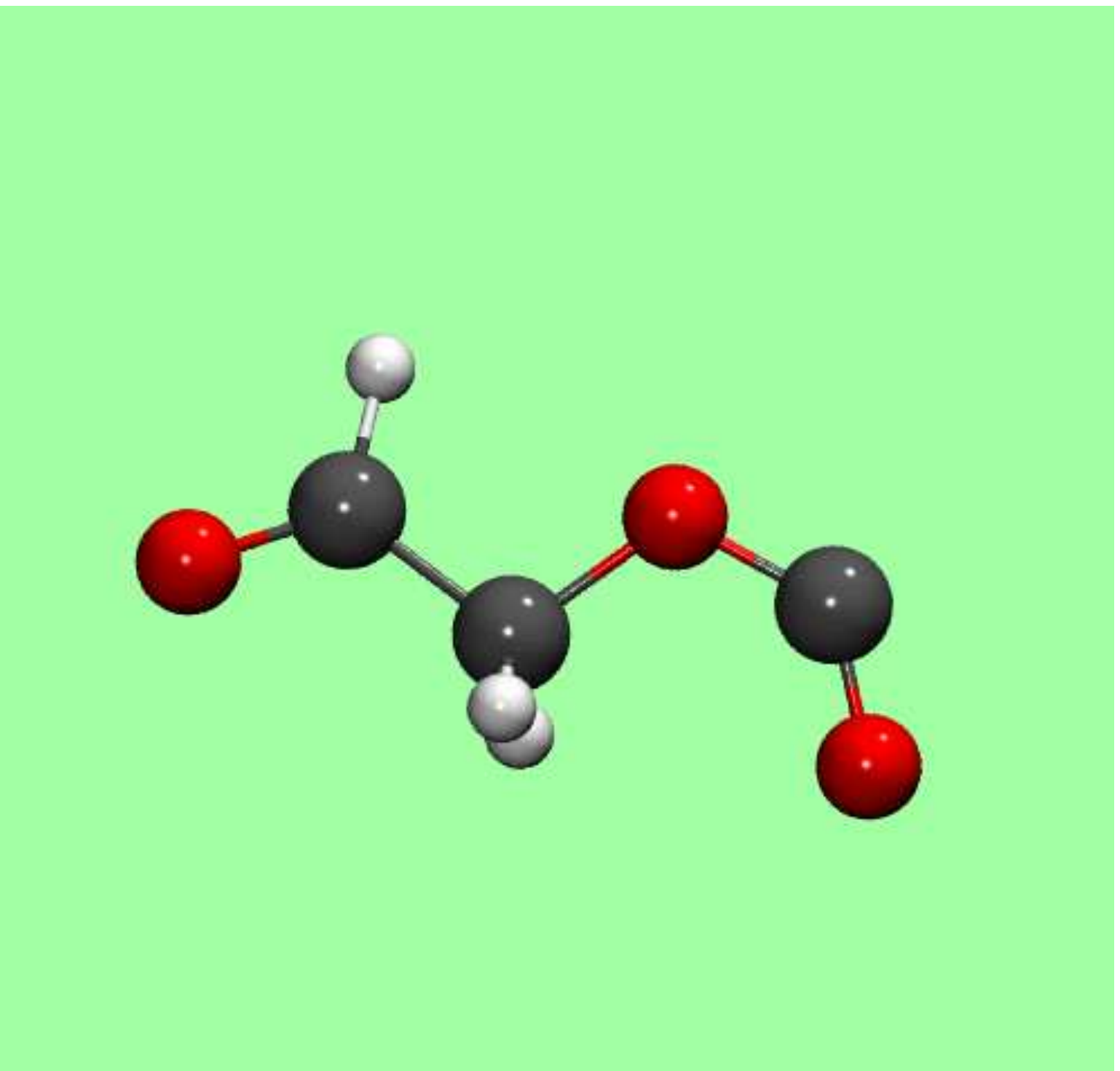}
		   \includegraphics*[width=2.15in]{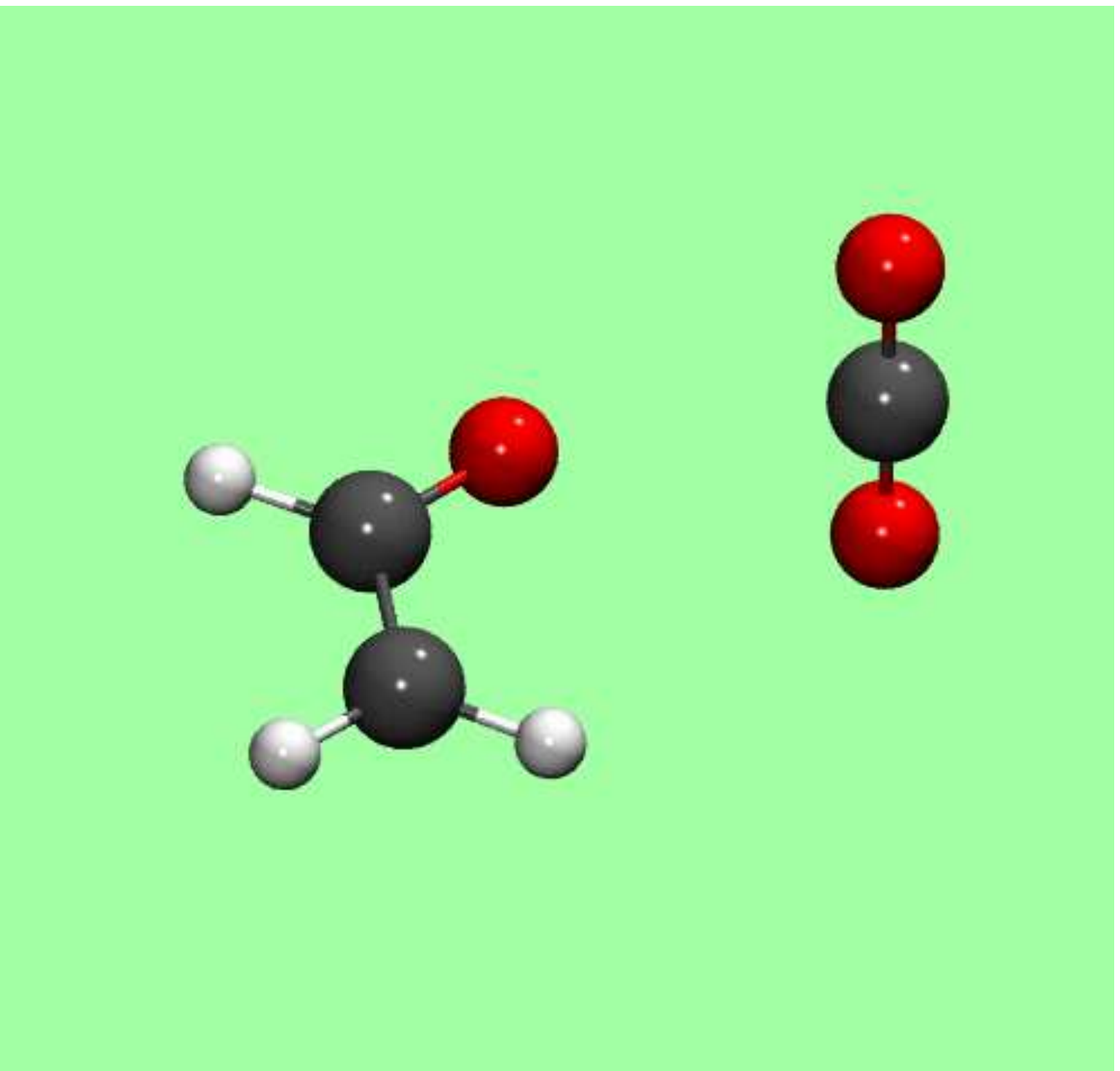} }}
\centerline{\hbox{ (a):G \hspace*{1.85in} (b):H \hspace*{1.85in} (c):I}}
\centerline{\hbox{ \includegraphics*[width=2.15in]{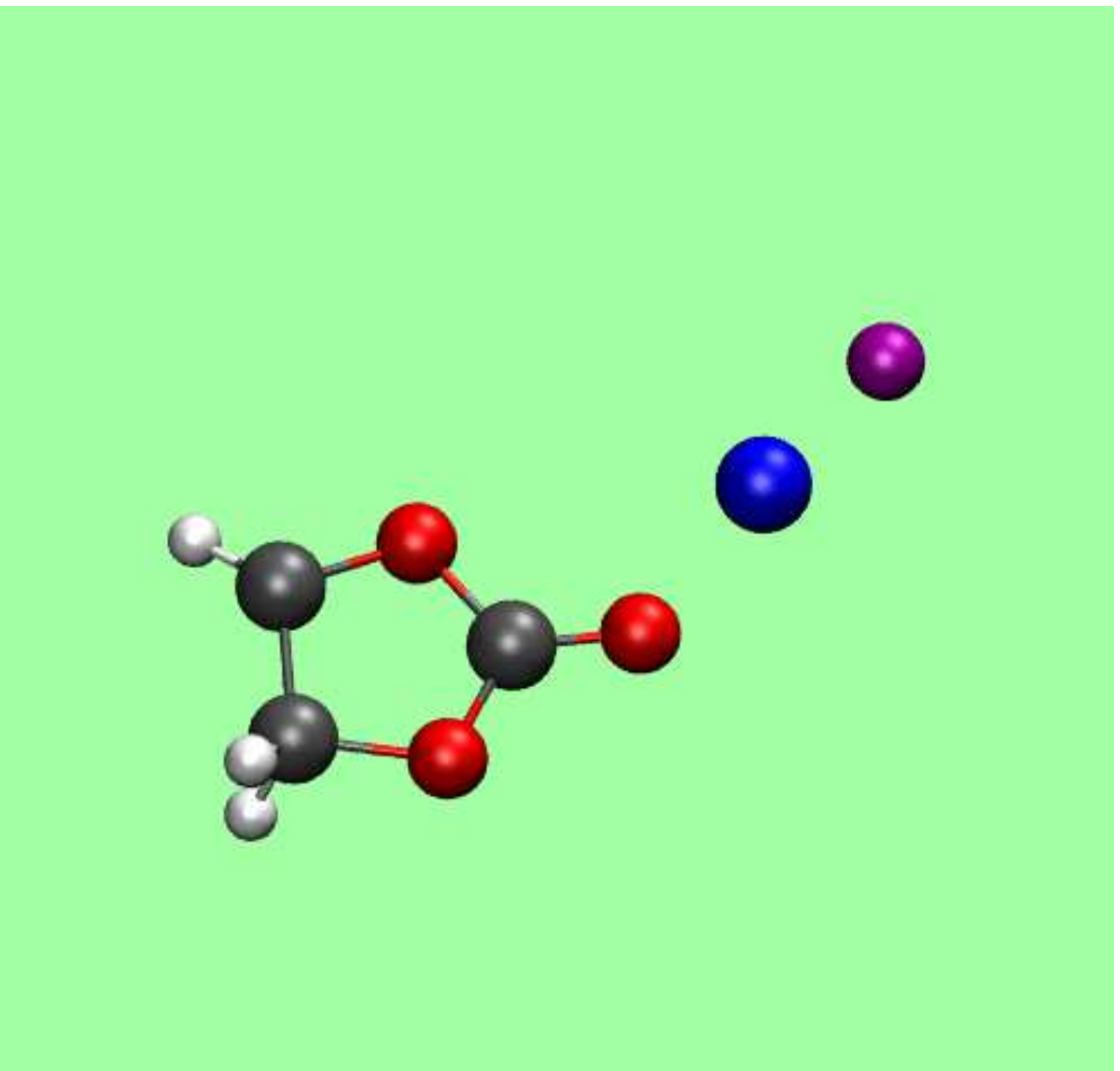}
		   \includegraphics*[width=2.35in]{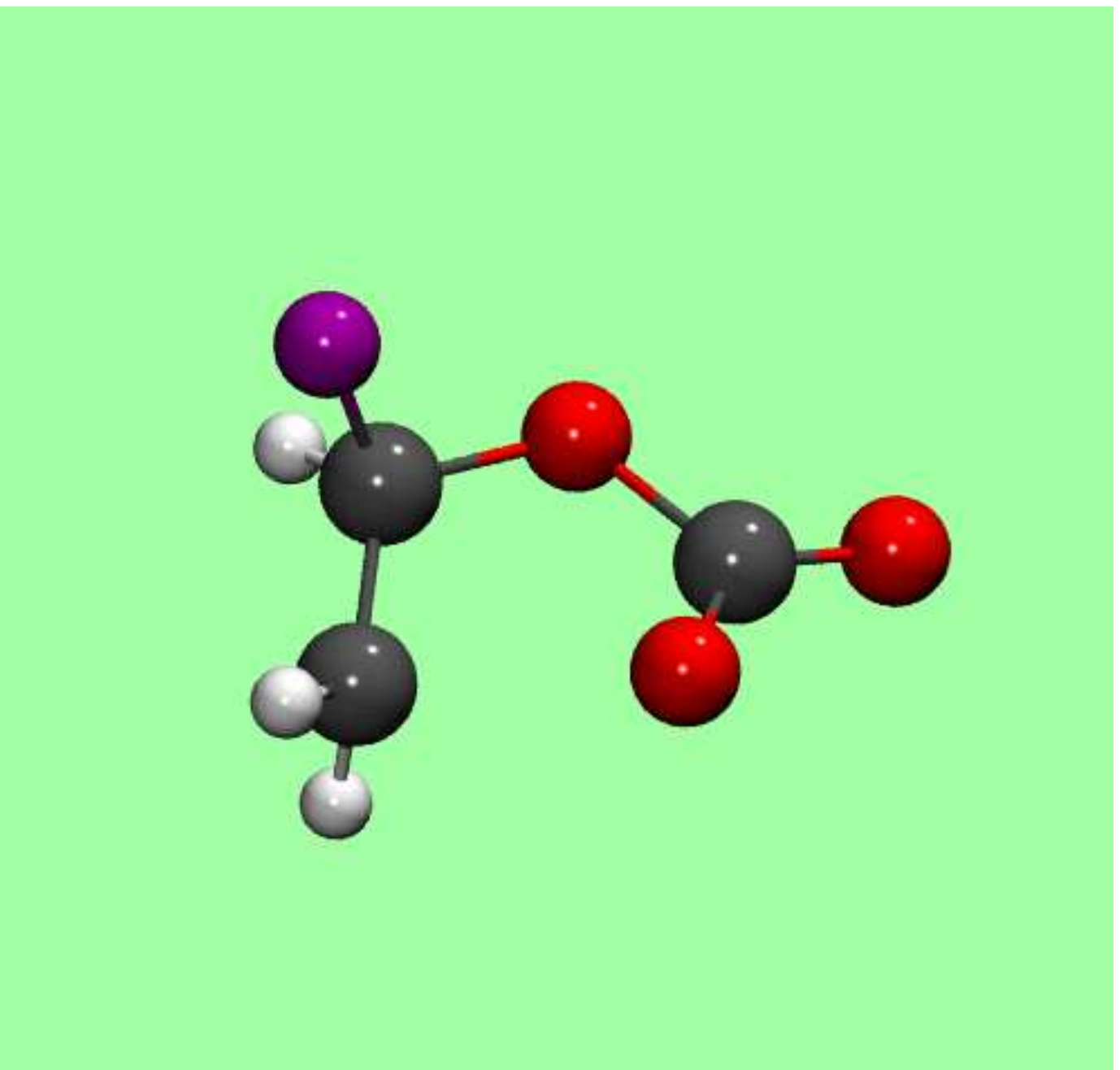}
		   \includegraphics*[width=2.15in]{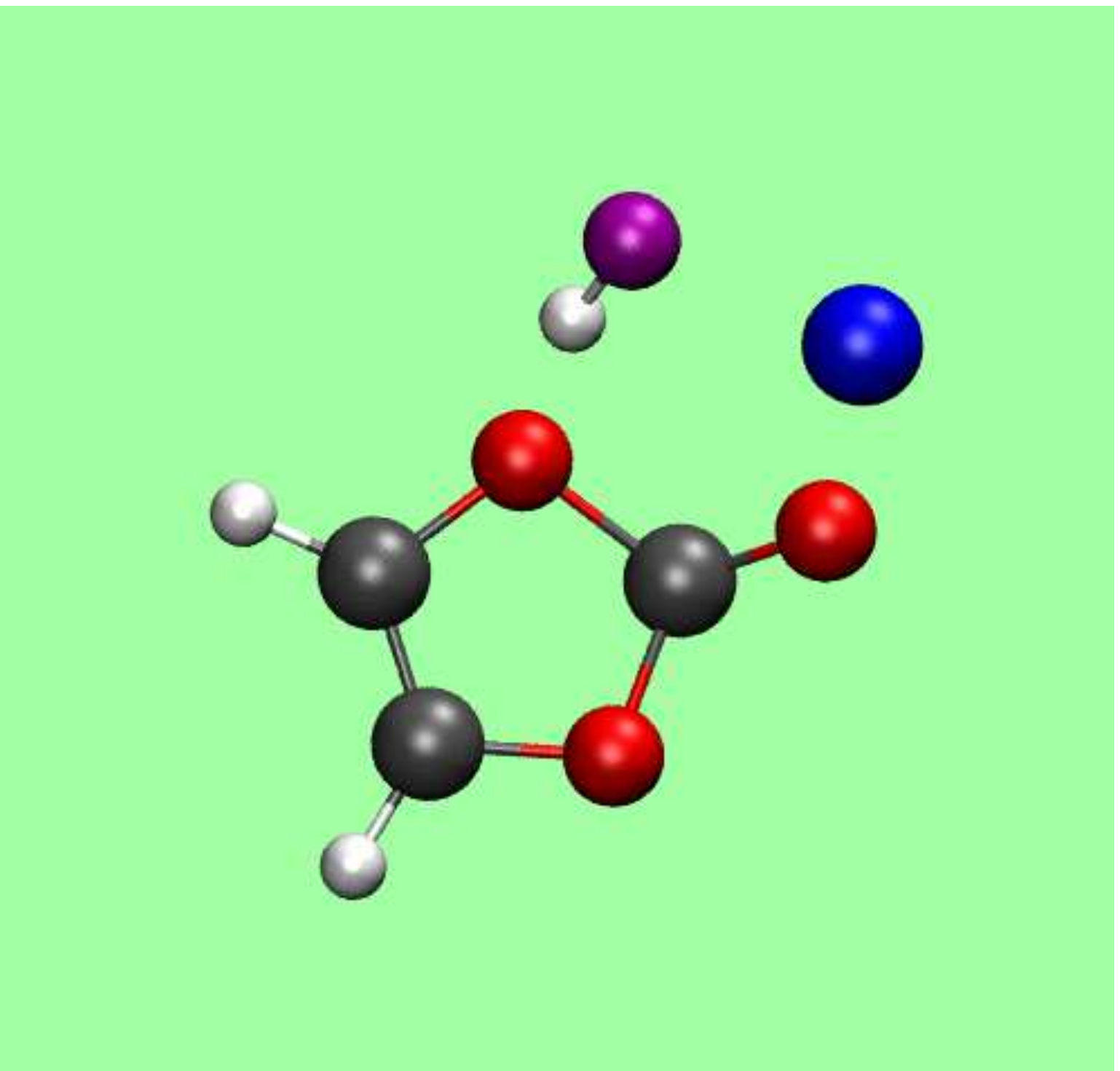} }}
\centerline{\hbox{ (d):J \hspace*{1.85in} (e):K \hspace*{1.85in} (f):L}}
\caption[]
{\label{fig4} \noindent
Configurations {\bf G}-{\bf L} depict broken FEC$^-$.  Panel (b) is
the same fragment as (a) after removing Li$^+$ and F$^-$ and re-optimizing.
The total energy of panel (d) is computed using two separate calculations
for CO$_2$ and $\cdot$CH$_2$CHO.  The most favorable pathway is 
{\bf A}$\rightarrow${\bf F}$\rightarrow${\bf G}/{\bf H}$\rightarrow${\bf I}.
Note that a 90$^o$ rotation about the C-C bond in {\bf H} is needed
in the transition state to eliminate CO$_2$.
}
\end{figure}

\begin{figure}
\centerline{\hbox{ \includegraphics*[width=5.in]{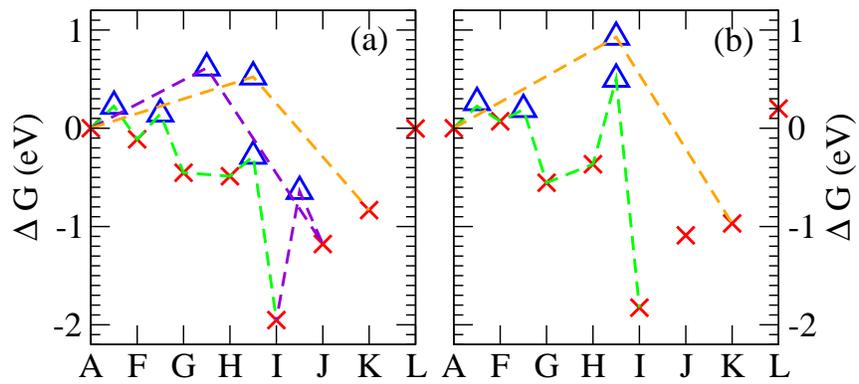} }}
\caption[]
{\label{fig5} \noindent
Free energies of intermediates (crosses) and barriers (triangles) associated
with configurations {\bf A}, {\bf F}, {\bf G}, {\bf H}, {\bf I}, {\bf J}, and
{\bf K}.  (a) PBE predictions; (b) MP2.  The path depicted in green is most
favored.  {\bf G} and {\bf H} differ by Li$^+$/F$^-$ diffusing away.
}
\end{figure}

Next we consider possible subsequent steps.
Guided by 2-$e^-$ reactions (Fig.~\ref{fig1}e), we next examine C-F bond
cleavage that proceeds from {\bf F}~to~{\bf G} (Fig.~\ref{fig4}a).  MP2
calculations reveal that the reaction is exothermic by 0.55~eV
(or 0.64~eV measured from the starting structure {\bf A}, Table~\ref{table1}).
The barrier from {\bf F}~to~{\bf G} is very small (0.12~eV,
Table~\ref{table2}).  (Using {\bf F} as reference implies that {\bf F}~is
sufficiently thermalized following bond-breaking reaction from~{\bf A}.)
The corresponding PBE values are 0.45~eV, 0.34~eV, and 0.26~eV, respectively.
Hence the same two-step reaction seen in a majority of two-electron reactions
(Fig.~\ref{fig1}b) is also thermodynamically and kinetically viable for
singly reduced FEC.  Compared with the AIMD simulations depicted in
Fig.~\ref{fig2}, the reason the C-F bond remains intact there seems to have
mostly entropic origins.  The lone Li$^+$ ion is far removed from
the F atom in Fig.~\ref{fig2}, and a longer trajectory appears necessary for
Li$^+$ to diffuse to and coordinate with the fluorine atom so the latter can
detach from the FEC$^-$ fragment.  

Due to the reactive nature of partially decomposed FEC$^-$, numerous product
channels and transition states exist for subsequent steps.  In some
cases the Li$^+$ and/or detached F$^-$ ions are omitted to assist calculation
of the specific barriers.  These models mimic conditions such that Li$^+$
and/or F$^-$ have ``diffused away.'' 

We next remove Li$^+$ and F$^-$ from {\bf G} and reoptimize the
geometry to {\bf H}.  The configuration shown in Fig.~\ref{fig4}b is the
most stable among several conformations separated by low rotational barriers.
Our attempt to initiate attack on either the C$_{\rm C}$ or C$_{\rm E}$ atom
of an intact EC molecule with the C or O atom of {\bf H} leads to endothermic
reactions with substantial barriers.  

Unimolecular reactions exhibit much
lower barriers and are much more exothermic.  Breaking the C$_{\rm E}$-O$_2$
bond of {\bf H} to form CO$_2$ and $\cdot$CH2CHO ({\bf I}, Fig.~\ref{fig4}c)
yields a MP2 (PBE) barrier of 0.869 (0.205)~eV and  the most
exothermic products found (Table~\ref{table1}, Fig.~\ref{fig5}).  
A large discrepancy between MP2 and non-hybrid DFT C$_{\rm E}$-O cleavage
barriers is observed.  A similar discrepancy has been documented for
the gas phase Li$^+$:EC$^-$ complex.\cite{han04}  Compared to
highly accurate CCSD(T) calculations, the barrier for breaking the
C$_{\rm E}$-O bond in EC$^-$ is overestimated by $\sim 0.15$~eV when using
MP2 and underestimated by BLYP by $\sim 0.33$~eV.\cite{han04} The hybrid
B3LYP DFT functional, which gives predictions closer to the CCSD(T) method
for EC$^-$, predicts a more modest 0.392eV barrier for breaking this bond
in FEC.  Even the likely overestimated MP2 C$_{\rm E}$-O cleavage barrier
does not preclude the unimolecular reaction from occurring during battery
charging time scales ($<$~1~hour) at room temperature. 

Note that the free energy of {\bf H}
is referenced to cluster {\bf A}, which has a different number of atoms,
via Configuration {\bf J} (Fig.~\ref{fig4}d to be discussed below).  {\bf J}
is optimized with and without Li$^+$ and F$^-$; its free energy is assumed
to be the same with or without these ions.  Using this procedure, {\bf G}
(with Li$^+$/F$^-$) and {\bf H} (without) are predicted to be similar
in MP2 (PBE) free energy, within 0.19~eV (0.03~eV) of each other.  This 
level of agreement is acceptable for our purposes.

Depending on the rate of $e^-$ transfer through the nascent SEI film and the
concentration of intermediates, the radicals {\bf F} and {\bf G} may
dimerize, or they may absorb extra $e^-$ and participate in further reactions,
like EC$^-$ and EC$^{2-}$.\cite{e2} Enumerating subsequent reactions that
may involve redox steps is beyond the scope of this work.  The implication of
proposed alternate fates of FEC on SEI formation is re-examined
in Sec.~\ref{discuss}.

CO elimination from {\bf H} is predicted to be endothermic by
1.28 (1.36)~eV when computed using the MP2 (PBE) method, and will not proceed.

\subsection{Cluster Calculations: Other Initiation Steps}

Again motivated by 2-$e^-$ reactions, we next consider the direct breaking
of the C-F in {\bf A} to form the charge-neutral radical {\bf J} with an
intact 5-atom ring.  According to PBE calculations, the reaction is exothermic
by 1.18~eV with a barrier of 0.614~eV.  {\bf J} can then undergo
C$_{\rm E}$-O$_2$ bond breaking to yield CO$_2$ and $\cdot$CH$_2$CHO ({\bf I},
discussed above; purple line in Fig.~\ref{fig5}).  This unimolecular reaction
that releases CO$_2$ is reminiscent of a VC-based reaction predicted in
Ref.~\onlinecite{okuno} for the reaction between VC and EC$^-$.  Unlike that
case, CO$_2$ is predicted to be a biproduct which accompanies the
release of uncharged organic radicals that may not form effective SEI
components.  The PBE barrier assocated with the first step this route 
is much higher than the one examined in the last subsection (green line),
and we have not computed the MP2 barriers associated with it.  The final
products are the same as in the last subsection.  {\bf J}
can also be obtained from ring-reformation from {\bf G}; it can 
can potentially undergo polymerization reactions.\cite{abraham}

We have also considered the cleavage of the C$_{\rm E}$-O$_2$ bond in {\bf A}
to form~{\bf K} (Fig.~\ref{fig4}e).  The barrier is found to be 0.927 (0.408)~eV
(orange line in Fig.~\ref{fig5}).  Unlike all other barriers considered
so far, {\it but} analogous to breaking the same bond in the ring-opened
radical (Table~\ref{table2}), the MP2 and PBE barriers differ significantly,
as noted earlier.  Even the likely
underestimated PBE barrier listed here gives a values higher than breaking
the C$_{\rm C}$-O bond ({\bf A}$\rightarrow${\bf F}), suggesting that
C$_{\rm E}$-O bond cleavage is kinetically unfavorable in FEC$^-$.
(See also appendix B.)  This {\bf A}$\rightarrow${\bf K} pathway is not
further considered, but it may still occur if aided by the electrode surface.
Breaking the C$_{\rm E}$-O$_1$ bond is less exothermic; the barrier has
not been computed because transition state searches have ended in the
transition state that yields C$_{\rm E}$-O$_2$ bond-breaking.

Elimination of HF from FEC$^-$ yields configuration~{\bf L} (Fig.~\ref{fig4}f)
which is endothermic by 0.20~eV (0.00~eV).  No low barrier pathway for
this two-bond-breaking reaction is found.

\subsection{Cluster Calculations: System Size Effect}

\begin{figure}
\centerline{\hbox{ (a):M \includegraphics*[width=2.15in]{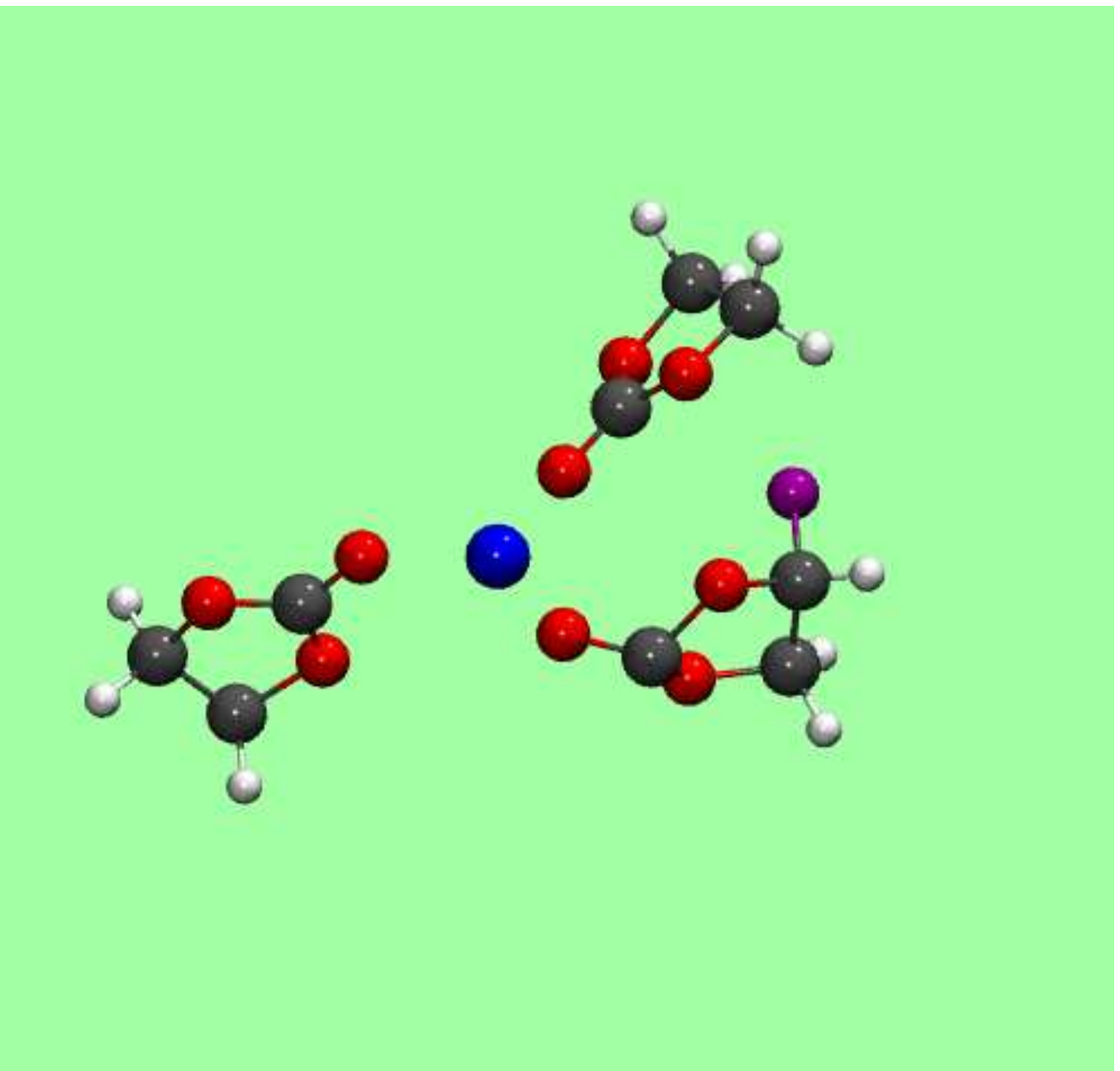}
		   \includegraphics*[width=2.15in]{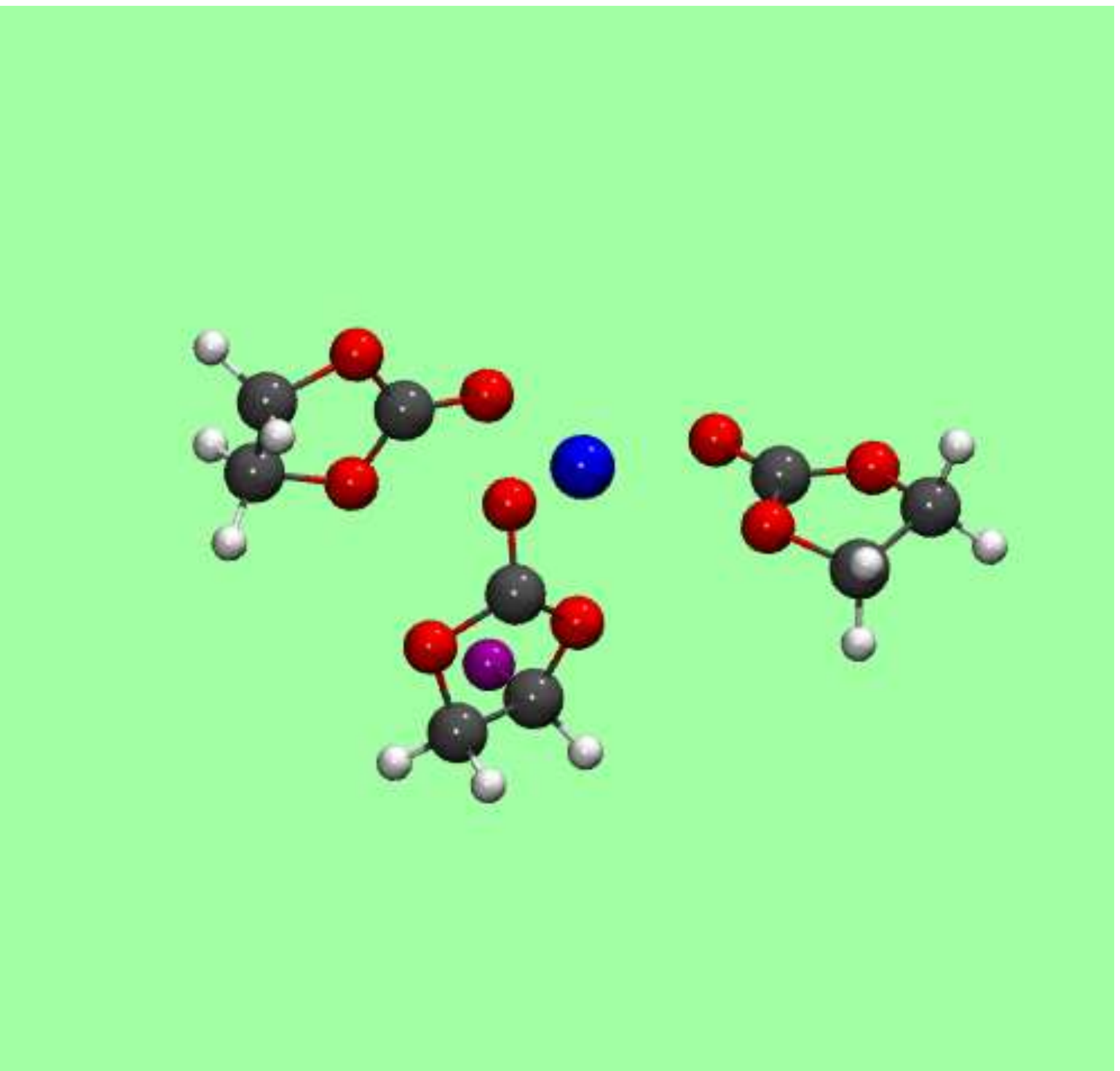} (b):N }}
\centerline{\hbox{ (c):O \includegraphics*[width=2.15in]{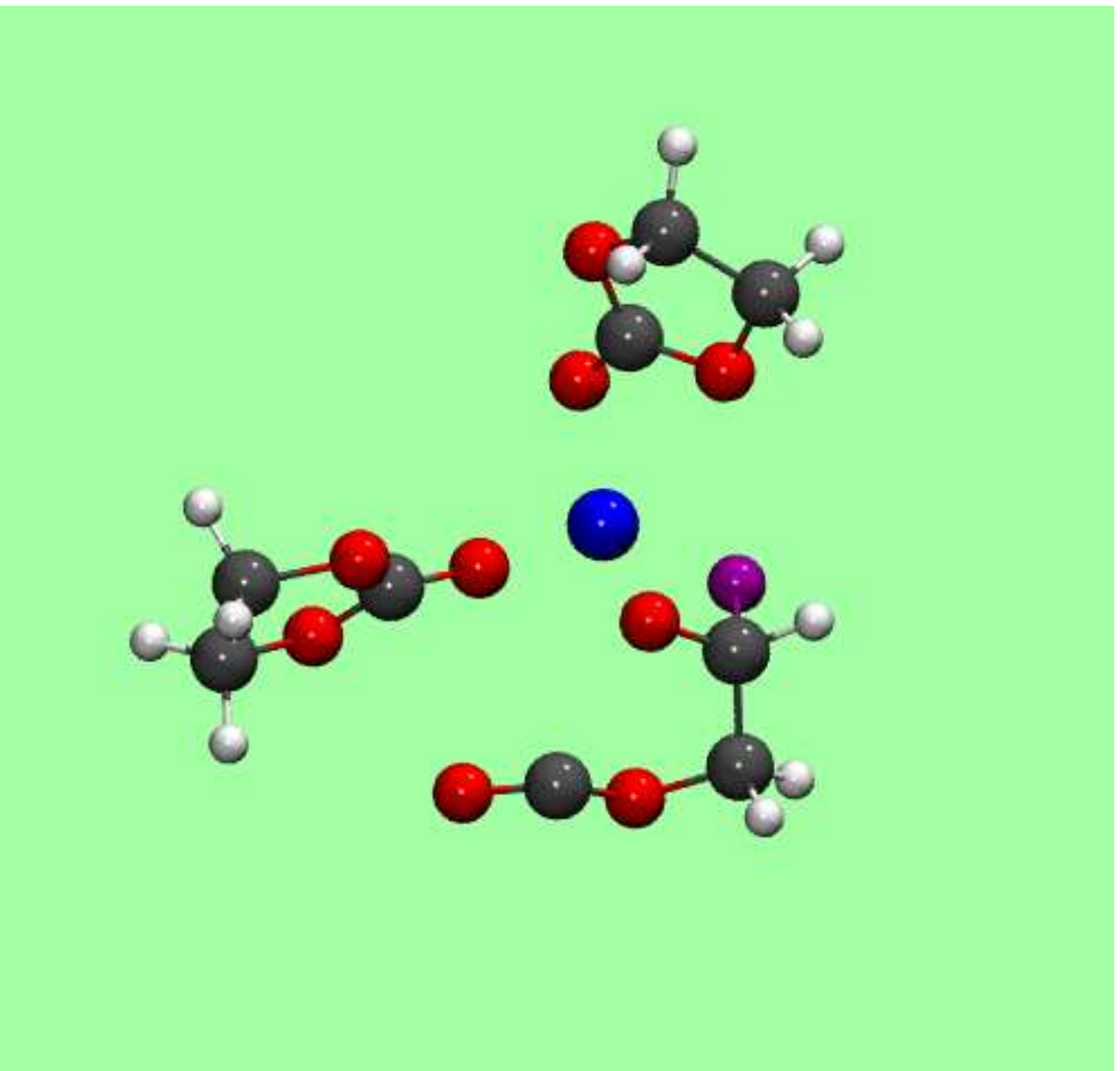}
		   \includegraphics*[width=2.15in]{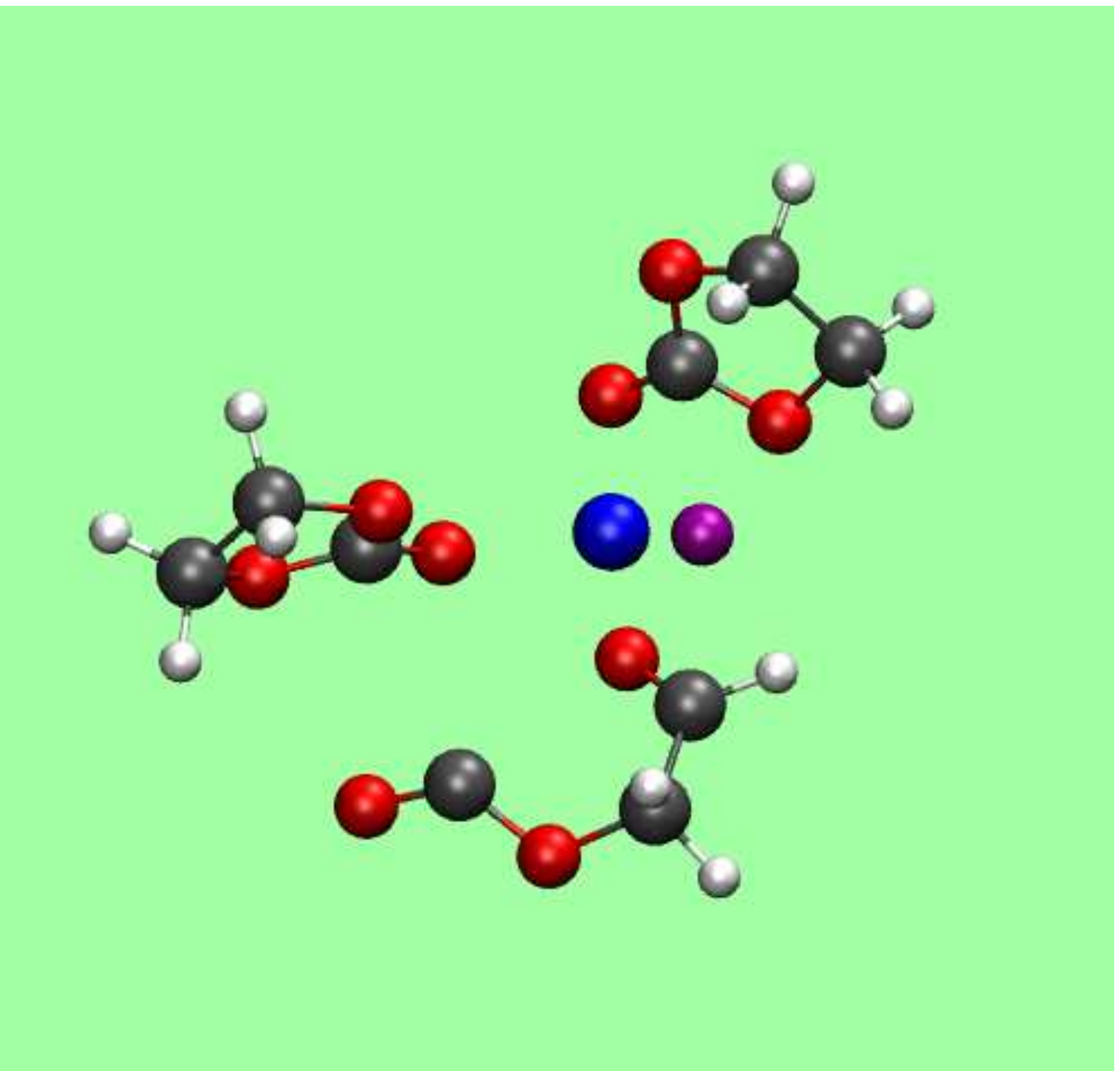} (d):P }}
\caption[]
{\label{fig6} \noindent
Configurations {\bf M}-{\bf P}, using Li$^+$:FEC$^-$(EC)$_2$ cluster models.
}
\end{figure}

So far, the results are derived using a cluster with one FEC, plus a
dielectric continuum that represents spectator solvent molecules.  In
experiments, the solvent is often a combination of EC and DMC, with FEC also
present at low concentration.  A more accurate way of representing the
FEC$^-$:Li$^+$ solvation shell is to introduce more molecules.
Configurations {\bf M}-{\bf P} (Figs.~\ref{fig6}a-d) depict optimized
geometries when two extra EC are included in the Li$^+$ coordination shell.
Using the PBE functional, the most favorable intact FEC configuration~{\bf M}
(Fig.~\ref{fig6}a) resembles the FEC$^-$:Li$^+$ cluster~{\bf E}
(Fig.~\ref{fig3}e), where Li$^+$ only coordinates to one FEC$^-$ O~atom.
{\bf N}~(Fig.~\ref{fig6}b) is analogous to {\bf A} (Fig.~\ref{fig3}a) and
is metastable, but by only 0.070~eV.  In contrast, MP2 predicts that {\bf M}
and {\bf N} are almost isoenergetic, emphasizing the somewhat sensitive
dependence of their free energy ordering on computational methods and cluster
size.  Configuration~{\bf O}
with a broken C$_{\rm C}$-O bond (Fig.~\ref{fig6}c) is analogous to
configuration~{\bf F}, but exhibits a lower exothermocity.  {\bf P}
(Fig.~\ref{fig6}d) is similar to {\bf G} (Fig.~\ref{fig4}c) with a smaller
release of free energy.  {\bf O} cannot be optimized using the MP2 method;
it decomposes into {\bf P} spontaneously.

We have not computed barriers using these larger clusters.  Recall, however,
that a C$_{\rm C}$-O$_1$ bond breaks within 5.3~ps in AIMD/PBE simulations
that treat {\it all} solvation molecules explicitly (Fig.~\ref{fig2}).
This timescale is consistent with at most a 0.1~eV barrier.  Thus,
depending on the electronic structure method, cluster size, and treatment
of temperature and out-lying spectator EC molecules, C$_{\rm C}$-O$_1$
bond breaking barriers vary from $\sim 0.1$~eV to a maximum of 0.26~eV.
Such estimated barriers are consistent with FEC$^-$ lifetimes that are
much shorter than battery charge/discharge timescales.

The main finding of this subsection is that using the smaller Li$^+$:FEC$^-$
complex may slightly overestimate reaction exothermicities.

\section{Discussions}
\label{discuss}

At least two requirements must be satisfied for good electrolyte additives: 1)
they should react at voltages higher than electrolyte solvents/anions
to ensure the SEI film is dominated by their breakdown products; 2) their
reaction intermediates must also coalesce and/or further react to form
SEI films that effectively transmit Li$^+$ ions but block solvent and
electron transport.  Li$^+$-containing ionic solids made up of negatively
charged species from additive/electrolyte decomposition are likely crucial for
Li$^+$ transport.  SEI made of purely charge-neutral oligomer products
which do not contain Li$^+$ are unlikely to conduct Li$^+$ efficiently,
or for that matter become dense/compact enough to block solvent transport.
Hence, both prediction of reduction potentials and elucidation of 
decomposition pathways are needed to define computational screening criteria
that lead to design of better additives.\cite{tasaki2010}

From this perspective, two-electron products (F$^-$ and C$_3$H$_3$O$_3^-$)
appear reasonable SEI components, although the latter contains X-CHO$^-$
groups that will likely undergo further reactions.  Dimerization is observed;
subsequent oligomerization may be possible.  The most favorable sequence
of one-electron reactions yields F$^-$, uncharged radicals, and CO$_2$.
This suggests that the 1-e$^-$
route yields LiF, found in the SEI, and other uncharged organic products that,
if they coaleasce, are likely neither very conductive to Li$^+$ nor form
compact, effective SEI films.  However, it cannot be ruled out that CO$_2$
and CHOCH$_2$$\cdot$ can be further reduced electrochemically and then
become part of the SEI.  Indeed, CO$_2$ reduction may yield oxalate
(C$_2$O$_4$$^{2-}$), as reported in some measurements.\cite{fec1,fec3}  The
likelihood of these subsequent events should depend on the relative rates
of electron transfer through the nascent SEI film and of
preciptation/nucleation of charged fragments on the electrode surface. 

Thus both 1- and 2-$e^-$ mechanisms lead to rapid release of F$^-$ to form
LiF.  We propose this is the main consequence of using FEC as additive.  
LiF can also arise from decomposition of PF$_6^-$, which is often considered
not to be an electrochemical reduction, but is related to H$_2$O 
contamination.  Hence FEC may generate LiF much more quickly.

Polycarbonates, suggested in some of the FEC/Si-anode
literature,\cite{fec1,fec2,fec3,fec5} is not directly observed in
FEC$^{n-}$ decomposition.  A previous computational work by one of the
authors proposed polycarbonate chain formation from two-electron reduction
of ethylene carbonate (EC).  Such chains are predicted to be unstable in high
dielectric environments, their ROCO$_2$R' motifs as prone to electrochemical
reduction and C$_{\rm C}$-O and C$_{\rm E}$-O cleavage as EC itself.  However,
if polycarbonate chains form large aggregates, the dielectric constant may
be sufficiently lowered that they are no longer susceptible to excess $e^-$
attacks.  Thus polycarbonates cannot be ruled out even though the experimental
evidence for their existence does not appear conclusive.

We have also performed preliminary calculations of FEC$^-$ decomposition on
lithium silicate-covered electrode surfaces immersed in liquid electrolyte.
This mixed oxide is known to be present after cycling silicon anodes in
lithium ion batteries.  The thin ($\sim 7$~\AA) insulator oxide layers
intervening between the metallic Li$_{13}$Si$_4$ and the electrolyte yield
two-electron FEC decomposition reactions, and therefore appear not
sufficiently thick to perform the necessary function of slowing down
$e^-$ transfer sufficiently to limit FEC decomposition to 1-$e^-$ reactions.

We have used PBE, MP2, and in one case, B3LYP methods to compute reaction
barriers.  In most cases the predictions are similar, within $\sim$$0.1$~eV
of each other.  The exception is breaking of the C$_{\rm E}$-O bond,
where MP2(PBE) appears to overestimate (underestimate) the true
barrier (see also appendix B).

\section{Conclusions}

In conclusion, we have examined the decomposition mechanisms of
fluoroethylene carbonate (FEC), which is a promising additive
for improving passivating solid electrolyte interphase
(SEI) films on silicon anode surfaces in lithium ion batteries.
Various mechanisms are examined using unconstrained AIMD simulations
of explicit liquid electrolyte/Li$_{\rm 13}$S$_4$ interfaces and using
cluster-based calculations.  The two types of simulations are consistent
with two- and one-$e^-$ attack on FEC, respectively.

Multiple reaction products and pathways are observed in fast 2-$e^-$ induced
FEC breakdown.  The main products are carbon monoxide or its precursor,
F$^-$ anion, and negatively charged C- and O-containing fragments that
may undergo further reactions.  We also observe direct defluorination of
``FEC$^{2-}$,'' yielding a C$_3$H$_3$O$_3^-$ fragment with an intact 5-atom
ring which in turn initiates nucleophilic attack on a neighboring EC molecule.
F$^-$ can precipitate to form LiF, which has been shown to be abundantly present
in FEC-derived SEI films,\cite{fec0,fec1,fec2,fec3,fec4,fec5,fec6,fec7} while
the negatively charged organic fragments can also contribute to the
Li$^+$-conducting part of the SEI.

Similar initial bond-breaking routes and F$^-$ ion release are also found
during 1-$e^-$ induced reactions.  While reaction pathways and products
have not been exhaustively enumerated, the first reaction is predicted to
be the cleavage of the C$_{\rm C}$-O$_1$ bond regardless of whether
unbiased AIMD/PBE simulation or cluster-based MP2/PBE barrier calculation
is used.  This internal consistency suggests that the predictions are
reasonable.  The most probable single FEC$^-$ molecule products are predicted
to be F$^-$, CO$_2$, and CHOCH$_2$$\cdot$ radicals.  Future experiments can
validate this prediction by analyzing the gas product composition.  Neither
HF nor CHFCH$_2$, proposed in the literature,\cite{japan,vc} is found as
a favorable reaction product or intermediate.

\section*{Acknowledgement}

We thank Yang Liu and Louise Criscenti for useful discussions, and Yukihiro
Okuno for sharing Ref.~\onlinecite{okuno}.  Sandia National Laboratories is a
multiprogram laboratory managed and operated by Sandia Corporation, a wholly
owned subsidiary of Lockheed Martin Corporation, for the U.S.~Department of
Energy's National Nuclear Security Administration under contract
DE-AC04-94AL85000.  The work reported in the main text was supported
by Nanostructures for Electrical Energy Storage (NEES), an Energy Frontier
Research Center funded by the U.S.~Department of Energy, Office of Science,
Office of Basic Energy Sciences under Award Number DESC0001160. SB is funded
by Sandia LDRD program. YM, JMM, and
PBB were supported by the Assistant Secretary for Energy Efficiency
and Renewable Energy, Office of Vehicle Technologies of the U. S. Department of
Energy under Contract No. DE-AC02-05CH11231, Subcontract No. 7060634 under the
Batteries for Advanced Transportation Technologies (BATT) Program.
This research used resources of the National Energy
Research Scientific Computing Center, which is supported by the Office of
Science of the U.S. Department of Energy under Contract No. DE-AC02-05CH11231.


\section*{Appendix A: FEC on Li metal cluster}

\begin{figure}[h]
\centerline{\hbox{ (a) \includegraphics*[width=2.15in]{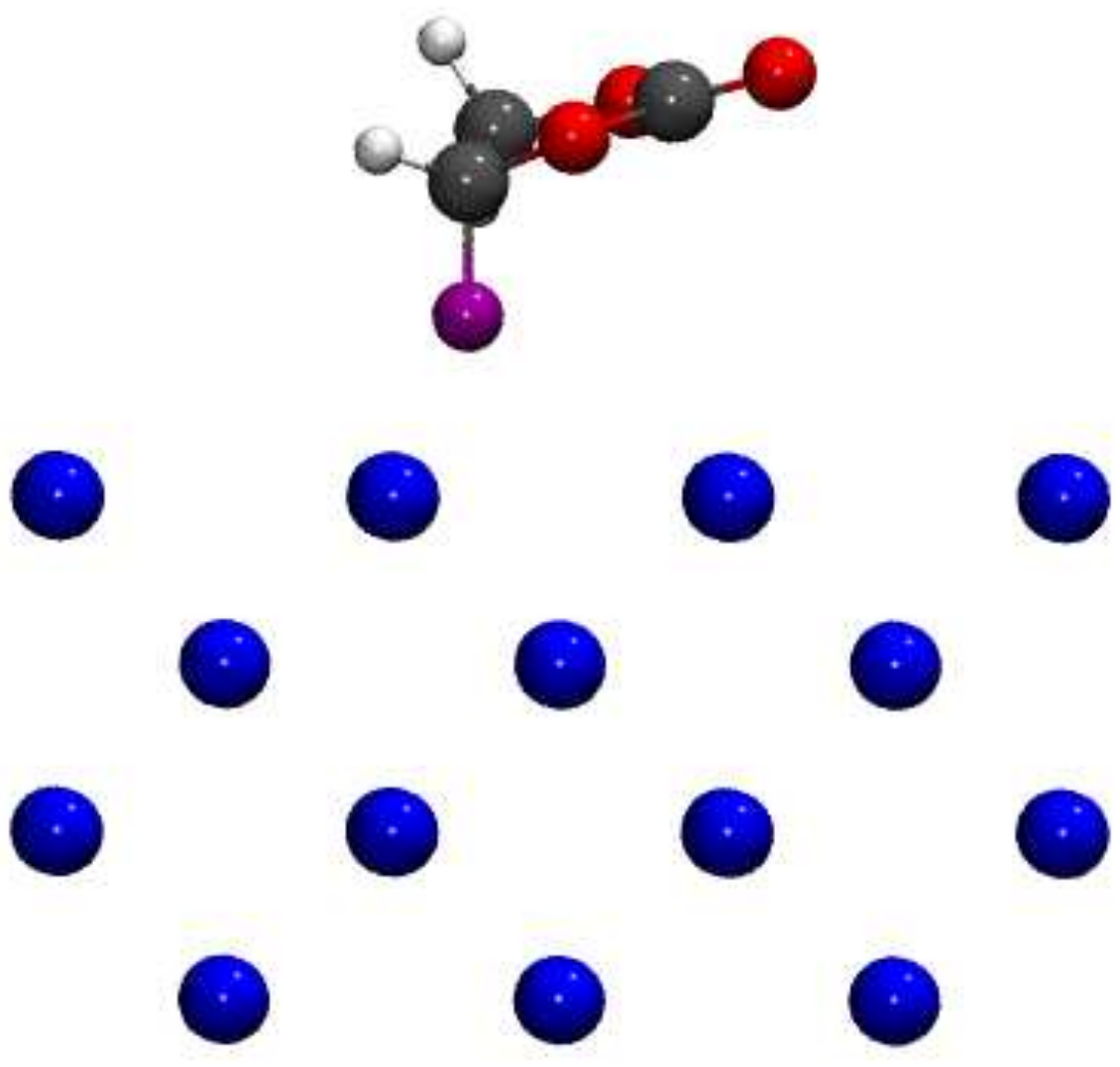}
		   \includegraphics*[width=2.15in]{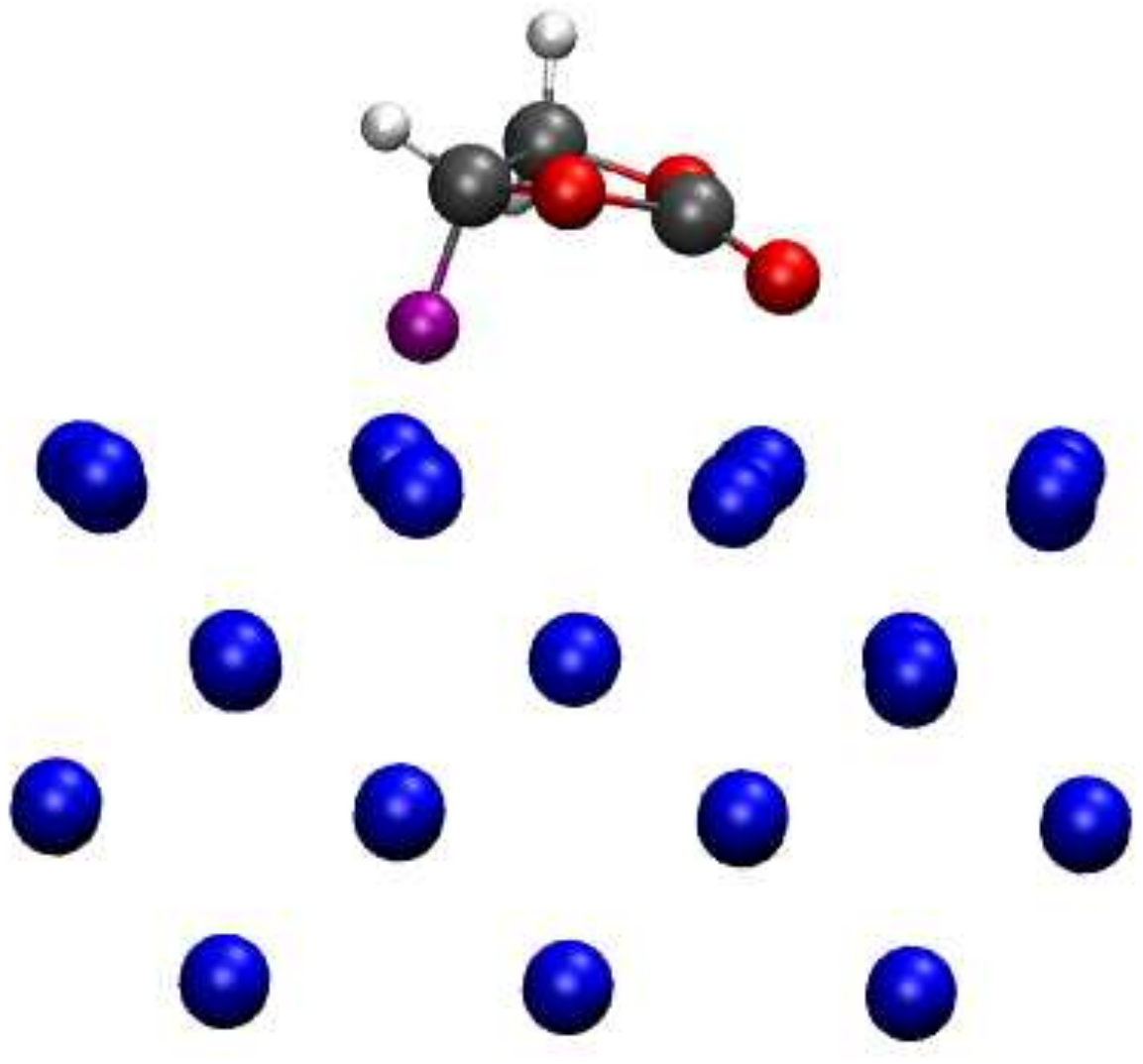} (b) }}
\centerline{\hbox{ (c) \includegraphics*[width=2.15in]{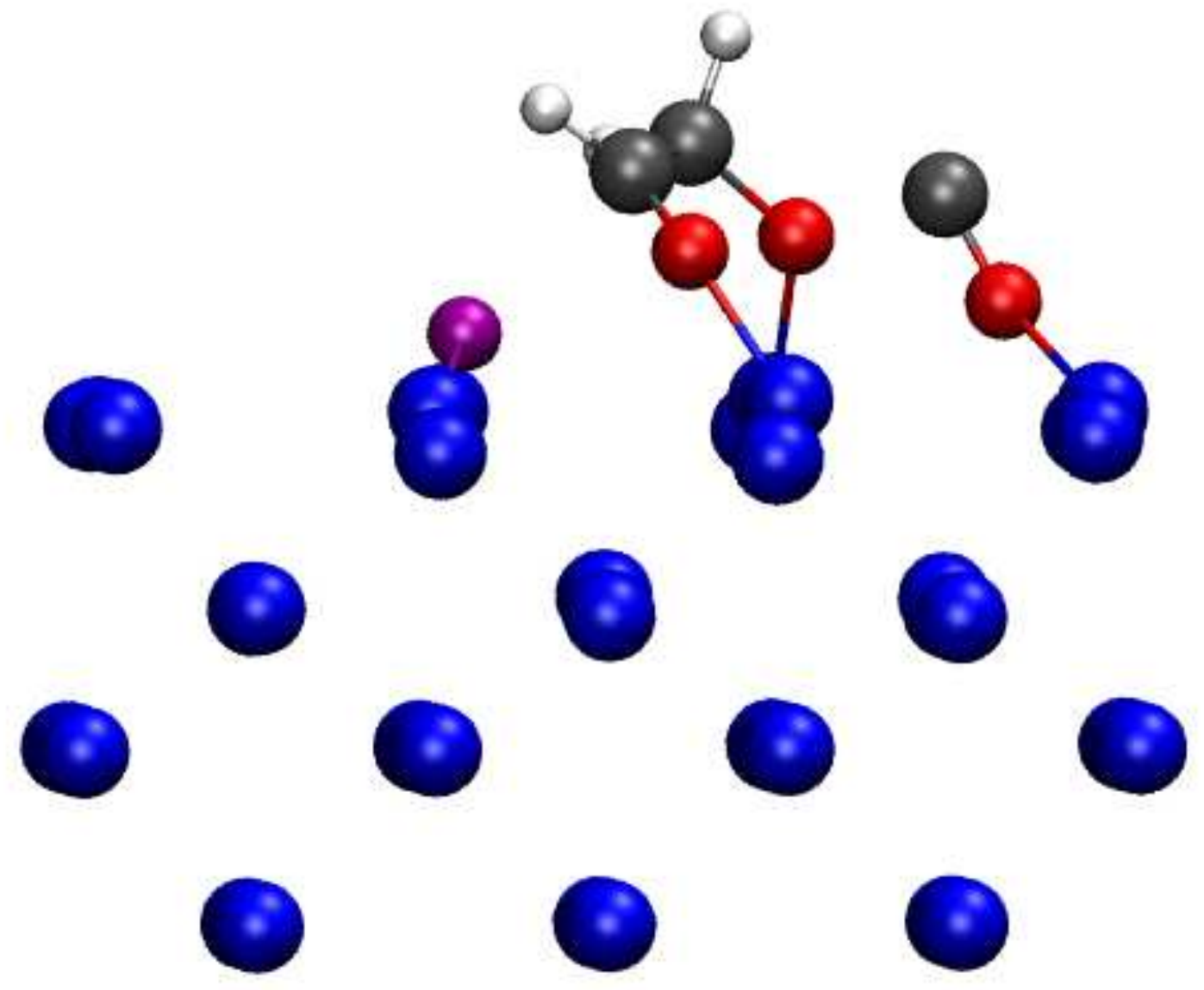}
		   \includegraphics*[width=2.15in]{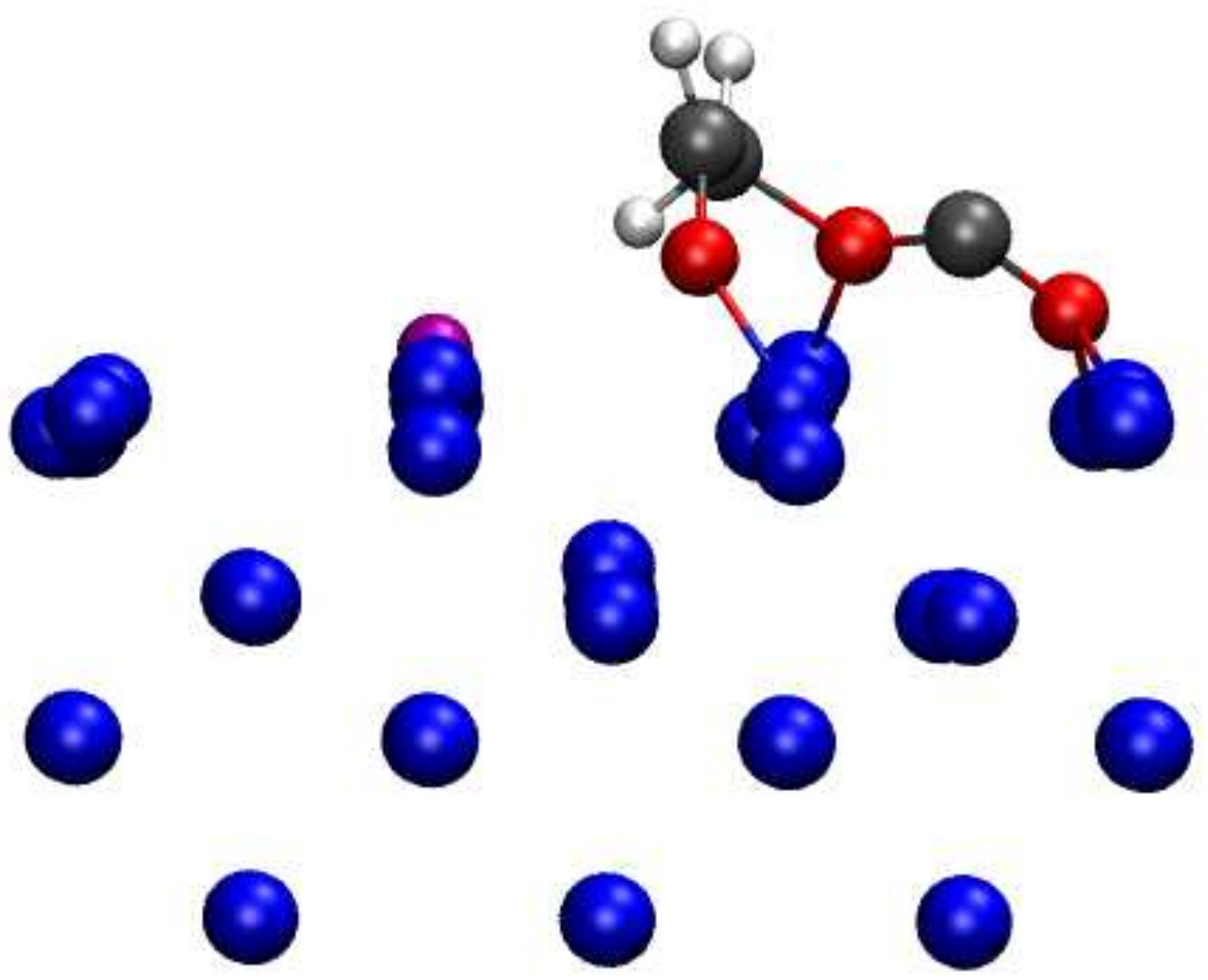} (d) }}
\caption[]
{\label{fig7} \noindent
Images of the decomposition of FEC on a Li surface as observed upon geometry
optimization at the B3LYP/{\tt 6-31G} level of theory using the ``smd''
dielectric continuum approximation.  The panels depict optimization
step number 0, 20, 40, and 100, respectively.
}
\end{figure}

To corroborate two-electron reduction pathways predicted in
Sec.~\ref{twoelectron} using hybrid DFT functionals,
which are generally more accurate than the DFT/PBE method applied in
the main text, an FEC molecule was placed on the (100) surface of a
50 atom face-centered cubic (FCC) Li cluster and optimized for 100 steps
at the B3LYP/{\tt 6-31G} level of theory\cite{becke,lyp} using the ``smd''
dielectric continuum approximation ($\epsilon$=40). The cluster consisted of
four layers of Li (Fig.~\ref{fig7}a).   The two bottom layers are frozen during
the optimization to help maintain the bulk FCC structure.

Upon optimization, the FEC molecule decomposes on the surface.  The initial
and intermediate structures are shown in Fig.~\ref{fig7}.  The C-F
bond stretches and the carbonyl oxygen migrates toward the surface (Step 20),
causing the C-O bonds to break and form carbon monoxide (Step 40).  The
F atom dissociates and binds to the Li surface.  These reaction features are
characteristic of two-$e^-$ attacks seen in DFT/PBE-based AIMD simulations
(Fig.~\ref{fig1}e).  In optimization step 80, the carbon monoxide fragment
recombines with one of the O atoms of the remaining FEC fragment.  In the
final step, a reduced ``CO$_2$
molecule'' forms, although the C$_{\rm E}$-O distance remains small (1.61~\AA)
and it may be argued that the C$_3$H$_3$O$_3$ fragment is one entity.
The last (100th) optimization step corresponds to the lowest energy structure
found, with a root-mean-square error Cartesian force of 0.003 Hartrees/Bohr.
F-containing solvent molecules for lithium air batteries have been predicted
to react with lithium clusters in a similar way, by breaking C-F bonds after
injection of electrons.\cite{liox}

\section*{Appendix B: FEC on Si-cluster}

A cluster model of Li$^+$:Si$_{15}$H$_{16}$ is employed to simulate the Si(100)
2$\times$1 surface and to examine FEC reduction reactions on the surface.
This model system might be relevant to silicon anodes with native oxide
films etched away by HF gas treatment.\cite{si_sei3}  The
calculations are performed using the Gaussian 09 suite of programs and
hybrid B3PW91 functional.\cite{becke,pw91} Previous studies have shown that
B3PW91 is reliable for the open-shell species involved in the ethylene
carbonate reduction reactions for lithium-ion batteries.\cite{bal01,bal02a}
Both geometry optimization and single-point calculations are carried out
using a {\tt 6-311++G(d,p)} basis set.  Thermal corrections are included
in the standard way, and analysis of atomic charges is generated using
the CHelpG method.\cite{chelpg} The adsorption energy $E_{\rm ads}$
is defined by $E_{\rm ads}$ = $E_{\rm sub/ads}$$-$$E_{\rm sub}$$-$$E_{\rm ads}$,
in which $E_{\rm sub/ads}$ is the total energy of the optimized
substrate-adsorbate structures, $E_{\rm sub}$ is the energy of the
substrate, and $E_{\rm ads}$ is the energy of adsorbate.
The Li$^+$:Si$_{15}$H$_{16}$ substrate will be discussed in detail
in a future publication.\cite{future}  No dielectric continuum approximation
is used in this appendix, and the reactions are therefore studied under
ultra-high vacuum conditions.

\begin{figure}
\centerline{\hbox{ \includegraphics*[width=5.in]{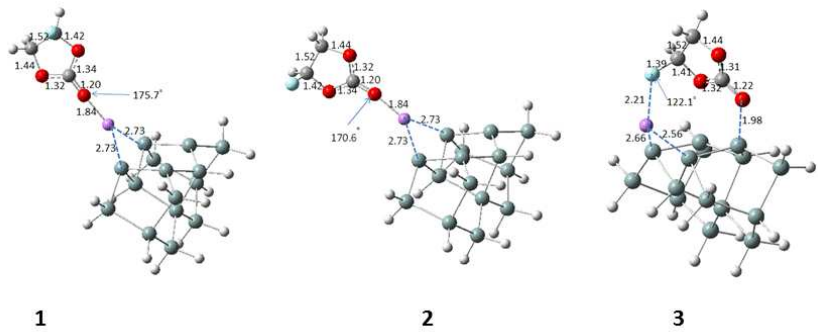} }}
\caption[]
{\label{fig8} \noindent
Calculated geometries for FEC adsorption on the Si$_{15}$H$_{16}$-Li$^+$
complex. C, O, H, F, Li and Si atoms are depicted as grey, red, white,
light blue, blue grey and purple spheres, respectively. Distances are in \AA.
}
\end{figure}

As discussed in the main text, both one-electron and two-electron mechanisms
are possible for decomposition of FEC.  We first examine FEC adsorption on the
optimized Si$_{15}$H$_{16}$-Li$^+$ cluster.  Three FEC adsorption modes are
identified, as illustrated in Fig.~\ref{fig8}. The major binding site is the
carbonyl O (O$_{\rm C}$) and the Li$^+$ in Configurations {\bf 1} and {\bf 2}.
{\bf 1} and {\bf 2} share similar geometries, and the only difference is the
position of the F atom: one is toward the Si(001) surface, whereas the other is
away from the surface. Not surprisingly, the DFT calculation indicates that
the two configurations possess approximately the same adsorption energy
($E_{\rm ads}$ = -1.14~eV).  Two binding sites are observed in
Configuration {\bf 3}: one is between the F atom and the Li$^+$, and the other
is between O$_{\rm C}$ and a surface Si atom. The FEC adsorption is slightly
weaker than that in {\bf 1} and {\bf 2}, with an adsorption energy of -0.93~eV,
suggesting that the O$_{\rm C}$-Li$^+$ interaction is stronger than F-Li$^+$.
The FEC geometry shows little change in the adsorbed structures.

The FEC-Li$^+$-Si$_{15}$H$_{16}$ cluster may receive one or two
electrons before decomposition takes place, depending on the
reduction potential, applied voltage, and electron transfer rate.  Our geometry
optimization produces 7 stable structures for the one-electron transfer
process. The most stable one, {\bf 4}, is obtained by adding an electron to and
optimizing from Configuration {\bf 2}, and the electron affinity (EA)  is
-5.46 eV.  The geometry of the FEC molecule is hardly changed
in the process, implying no significant electron structure change
of the adsorbed molecule.  This is further verified by charge distribution
analysis: most negative charge ($\sim$87\%) fluxes into the Si$_{15}$H$_{16}$
cluster, whereas the charge on the FEC molecule remains roughly
unchanged, as listed in Table~\ref{table4}.  Thus, without the dielectric
environment provided by the liquid electrolyte, the Si$_{15}$H$_{16}$
cluster retains most of the negative charge, retarding FEC decomposition.

Two-electron transfer reactions are also investigated and five stable
structures are obtained after geometry optimization. The energetically
most favorable structure is {\bf 5} generated by {\bf 1} + 2$e^-$
$\rightarrow$ {\bf 5} (combined first and second EA = -7.99 eV).  The
overall configuration of {\bf 5} differs substantially from {\bf 1}, in
which the FEC molecule is away from the cluster and the ring plane is
approximately perpendicular to the Si surface.  The ring resides on the top of
the cluster and is parallel to the Si(001) surface in Configuration {\bf 5}.
Another stable structure, {\bf 6}, is obtained using {\bf 3} as starting
configuration.  The combined 1st and 2nd EA is -8.10~eV for
{\bf 3} + 2 $e^-$ $\rightarrow${\bf 6}. In the electron transfer process,
O$_{\rm C}$ moves away from the Si surface. Therefore, only one major binding
site between the F-atom and the Li atoms is found in {\bf 6}.
Analogous to Configuration {\bf 4}, most negative charge transfers to the
Si$_{15}$H$_{16}$ cluster in {\bf 5} and {\bf 6}.  However, there is
significant ngative charge ($\sim$84\%) flowing into the FEC molecule.
For example, the net charge of FEC in {\bf 5} is $-0.15 |e|$ in
comparison to $0.07 |e|$ in 1.

Next we examine the first step of bond dissociation for both one- and
two-electron decomposition mechanisms. Five bond-breaking modes
A-E (not to be confused with configuration labels in the main text)
are illustrated in Fig.~\ref{fig9}. For one electron mechanisms, bond cleavage
occurs via A, B, C, D and B/D. DFT calculations fail to locate intermediates
and transition states for bond-breaking mode E.  Hence C-F breaking may
not be viable as an one-electron mechanism. The corresponding reaction
energies ($E_{\rm reac}$) and energy barriers ($E^*$) are listed in
Table~\ref{table5}.  $E_{\rm reac}$ vary between -0.48~eV and
0.68~eV. Bond cleavages via A, D, and B/D are exothermic whereas
B and C are endothermic.
This indicates that C$_{\rm E}$-O bond-breakings are thermodynamically
favorable for the one-electron mechanism. However, they are not kinetically
favorable due to the relatively high energy barrier (1.14-1.57~eV). The
lowest energy barrier is C$_{\rm c}$-O cleavage (mode C), with a value of
0.43~eV.  The intermediates usually have more binding site(s) on
the Si surface, primarily via the O and C$_{\rm E}$ atoms.

Four bond-breaking modes, A/D, B, C and E are identified for two electron
mechanisms (Table~\ref{table6}.  The reactions are more exothermic
($-1.86$ -- $-2.85$~eV) than those in one electron reactions.
Although bond-breaking mode A/D exhibits the most negative reaction energy,
its high energy barrier (1.24~eV) makes the bond cleavage kinetically
less favorable. One the other hand, bond cleavages at B and C show
very low energy barriers (0.11 and 0.28~eV). Therefore, C$_{\rm C}$-O
bond breakings via the two-electron mechanism are both thermodynamically
and kinetically favorable on the Li-Si$_{15}$H$_{16}$ cluster. This is in
full agreement with the AIMD calculations. Bond breaking at B leads to
the formation of intermediate {\bf 7}, in which 3 binding sites, O$_{\rm C}$-Li,
C$_{\rm C}$-Si, and O$_{\rm E}$-C, are observed (Fig.~\ref{fig10}).  The
multi-bindings stabilize the intermediate, and therefore may facilitate the
reduction process. The corresponding low-barrier transition state, TG, is
associated with formation of a C$_{\rm C}$-Si bond and cleavage of a
C$_{\rm C}$-O bond as well.  Furthermore, C-F bond breaking is found
to be feasible with two-electron reduction.  In intermediate {\bf 8}, a Li-F
bond is formed on the Si surface, and the ring structure is also adsorbed
on the surface (Fig~\ref{fig10}). Its transition state involves the
dissociation of C-F and the formation of Li-F, with a barrier of
0.72~eV. C-F bond breaking is also found in AIMD simulations.
In summary, two electron mechanisms are favorable for FEC reduction
on the Li-absorbed Si cluster while one electron mechanisms are either
endothermic or exhibit high barriers.  Whether one- or two-electron
routes dominate depends on electron transfer processes, which are beyond the
scope of this study.

\newpage

\begin{references}
\bibitem{note0}
Even the formation of SEI films from unsubstituted ethylene carbonate (EC)
molecules and cosolvent diethyl carbonate (DEC), used as standard laboratory
electrolyte, has arguably not been completely understood.\cite{e2,abraham}
\bibitem{insulating}
Since similar organic functional groups are identified on Si and graphite
anode surfaces, the pertinent SEI formation mechanisms is likely similar for
both anode materials.  One ready explanation is the cracking of Li$_x$Si$_y$
particles exposing pristine surfaces to allow 2-electron reduction.
\end{references}
\end{document}